\documentclass[a4paper,fleqn,usenatbib]{mnras}

\usepackage[T1]{fontenc}
\usepackage{ae,aecompl}

\usepackage{subfig}
\usepackage{relsize}

\usepackage{graphicx}	
\usepackage{amsmath}	
\usepackage{amssymb}	

\usepackage{booktabs}
\usepackage{longtable}
\usepackage{array}
\usepackage{pdflscape}

\newcommand{\eqb}{\begin{equation}}
\newcommand{\eqe}{\end{equation}}
\newcommand{\dmb}{\begin{displaymath}}
\newcommand{\dme}{\end{displaymath}}

\newcommand{\eab}{\begin{eqnarray}}
\newcommand{\eae}{\end{eqnarray}}

\newcommand{\e}{\mbox{e}}
\newcommand{\be}{\begin{equation}}
\newcommand{\ee}{\end{equation}}

\newcolumntype{C}{>{$\displaystyle} c <{$}}

\title[SU(2)$_{\text{CMB}}$ at high redshifts]{SU(2)$_{\text{CMB}}$ at high redshifts and the value of $H_0$}

\author[Hahn \& Hofmann]{
Steffen Hahn$^{1}$,
Ralf Hofmann$^{2}$
\\
$^{1}$Karlsruhe Institute of Technology, Laboratory for Applications of Synchrotron Radiation,  Kaiserstr. 12, Karlsruhe D-76131, Germany \\
$^{2}$Universit\"at Heidelberg, Institut f\"ur Theoretische Physik, Philosophenweg 16, Heidelberg D-69120, Germany
}

\date{Accepted XXX. Received YYY; in original form ZZZ}

\pubyear{2016}

\begin{document}
\label{firstpage}
\pagerange{\pageref{firstpage}--\pageref{lastpage}}
\maketitle

\begin{abstract}
We investigate a high-$z$ cosmological model to compute the co-moving 
sound horizon $r_s$ at baryon-velocity freeze-out towards the end of hydrogen 
recombination. This model 
assumes a replacement of the conventional CMB photon gas by deconfining SU(2) Yang-Mills thermodynamics, 
three flavours of massless neutrinos ($N_\nu=3$), and a {\sl purely baryonic} matter sector 
(no cold dark-matter (CDM)). The according SU(2) temperature-redshift relation of the 
CMB is contrasted with recent measurements appealing to 
the thermal Sunyaev-Zel'dovich effect and CMB-photon absorption by molecular rotations bands or atomic 
hyperfine levels. Relying on a realistic simulation of the ionization history throughout 
recombination, we obtain $z_*=1693.55\pm 6.98$ and $z_{\rm drag}=1812.66\pm 7.01$. Due to considerable widths 
of the visibility functions in the solutions to the associated Boltzmann hierarchy and Euler 
equation we conclude that $z_*$ and $z_{\rm drag}$ over-estimate the redshifts for the 
respective photon and baryon-velocity freeze-out. Realistic decoupling values turn out to be 
$z_{{\rm lf},*}=1554.89\pm 5.18$ and $z_{\rm lf,drag}=1659.30\pm 5.48$. 
With $r_s(z_{\rm lf,drag})=(137.19\pm 0.45)\,$Mpc and the essentially 
model independent extraction of $r_s\cdot H_0=\mbox{const}$ from low-$z$ data 
in arXiv:1607.05617 we obtain a good match with the value 
$H_0=(73.24\pm 1.74)\,$km\,s$^{-1}$\,Mpc$^{-1}$ extracted in 
arXiv:1604.01424 by appealing to Cepheid calibrated SNe~Ia, 
new parallax measurements, stronger constraints
on the Hubble flow, and a refined computation of distance to NGC4258 from maser data. 
We briefly comment on a possible interpolation of our 
high-$z$ model, invoking percolated and unpercolated U(1) topological solitons of a
 Planck-scale axion field, to the phenomenologically successful low-$z$ $\Lambda$CDM cosmology. 
\end{abstract}




\section{Introduction}

The last two and a half decades have witnessed a tremendous industry in collecting and interpreting precise observational 
data to determine the cosmology of our universe: (i) large-scale structure surveys confirming the existence of a standard 
ruler $r_s$ set by the physics of baryonic acoustic oscillations throughout the epochs preceding and 
including the recombination of primordial helium and hydrogen, 
e.g. \cite{Abazajian2003,Adelman-McCarthy2008}, (ii) observations of the cosmic microwave background (CMB), confirming its black-body nature 
and revealing the CMB decoupling physics as well as associated primordial statistical properties of 
matter and hence temperature fluctuations, e.g. \cite{Mather1990,Hinshaw2013,Ade2014a}, and (iii) 
use of calibrated standard candles in luminosity distance-redshift observations, ultimately 
changing the paradigm on late-time expansion history (low-$z$ regime) \cite{Perlmutter1998,Riess1998}. 
As a consequence, we now appear to possess an accurate parametrisation of 
the universe's composition in terms of the standard $\Lambda$CDM concordance model. 
Yet, we suspect that this model is prone to over-simplification: So far 
there is no falsifiable theory on what the dark sector actually is made of. Moreover, as we shall argue in the present work, the 
extrapolation of the observationally well established low-$z$ model to thermal expansion history well before and including CMB decoupling, 
although seemingly in accord with the results of \cite{Hinshaw2013,Ade2014a}, can be misleading. 

\begin{figure}
\centering
\includegraphics[width=\columnwidth]{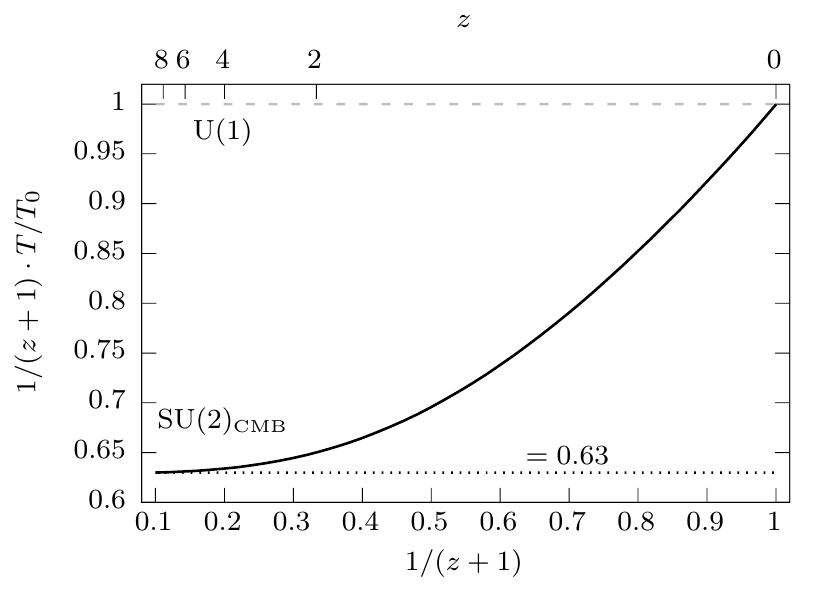}
\caption{\protect{\label{Fig-1}} The $T-z$ scaling relation ${T/(T_0(z+1))}$ in SU(2)$_{\text{CMB}}$ (solid). 
Note the emergence of ${T/T_0}=0.63(z+1)$ for $z\gtrsim 9$ (dotted). 
The conventional U(1) theory for thermal photon gases associates with 
the dashed line. Data taken from \protect\cite{Hofmann2015} after 
slight and inessential correction.}      
\end{figure}

In \cite{Hofmann2015} the implication of a new SU(2) Yang-Mills
theory, describing the CMB as a gas of thermal photons supplemented by a thermal ground state and two invisible vector mode $V^\pm$, towards the temperature-redshift ($T$-$z$) 
relation was analysed within an Friedmann-Lemaitre-Robertson-Walker (FLRW) universe. 
Due to non-conformal scaling at low $z$ this relation suffers a lower 
linear slope at high $z$ ($z\gtrsim 9$) compared to the standard U(1) theory\footnote{In Eq.\,(\ref{T-z}) a slight and inessential correction of the high-$z$ 
coefficient from 0.62 in \cite{Hofmann2015} to 0.63 is performed which is due to a more 
precise initial-condition matching for the solution to the energy-conservation equation.}, 
\begin{align}
\frac{T}{T_0}&=z+1\,,\ \ \ (\text{U(1)}, \forall z) \ \ \longrightarrow \nonumber \\
\frac{T}{T_0}&=0.63\,(z+1)\,,\ \ \ (\text{SU(2)}_{\text{CMB}}, z\gtrsim 9)\,,
\label{T-z}
\end{align}  
see also Fig.\,\ref{Fig-1}. We refrain here from reviewing the entire thermodynamics 
of an SU(2) Yang-Mills theory in its deconfining phase. We also skip a discussion of why the 
critical temperature $T_c=T_0$ ($T_0=2.725\,$K referring to the present CMB temperature) 
for the onset of the 
deconfining-preconfining phase transition in such a theory is fixed by low-frequency 
observation of the present CMB, justifying the name 
SU(2)$_{\text{CMB}}$. To be 
informed about all this in a pragmatic way we refer the reader to \cite{Hofmann2015}, 
for an in-depth read we propose \cite{Hofmann2016a}. We do mention though 
that the standard relation between neutrino temperature $T_\nu$ and $T$, $T_\nu=\left({4/11}\right)^{1/3}T$, 
obtained from entropy conservation during $e^+ e^-$ annihilation, modifies in 
SU(2)$_{\text{CMB}}$ to 
\eqb\label{TneutTcmbSU2}
\frac{T_\nu}{T}=\left(\frac{g_1}{g_0}\right)^{1/3}=\left(\frac{16}{23}\right)^{1/3}\,.
\eqe
This is because $g_0=8+({7/8})\,4$ (number of relativistic degrees of freedom (d.o.f.) before annihilation) and 
$g_1=8$ (d.o.f. after annihilation), see \cite{Hofmann2015}. Note that for, say, 
$z>100$ SU(2)$_{\text{CMB}}$'s two massive vector excitations $V^\pm$ can be considered 
highly relativistic. Namely, for $z>100$ one has ${m_{V^\pm}/T}<1.4\times 10^{-3}$ such 
that the energy density $\rho_{\text{SU(2)}_{\text{CMB}}}$ is due to eight 
relativistic d.o.f.  

The SU(2)$_{\text{CMB}}$ relation (\ref{T-z}) and Fig.\,\ref{Fig-1} represent a strong deviation from the conventional relation 
$\frac{T}{T_0}=z+1$, the latter being a direct consequence of the conformal, thermal photon-gas equation of 
state $p_\gamma=\frac13\rho_\gamma$. In SU(2)$_{\text{CMB}}$, however, this equation of state is non-conformal 
because it incorporates the thermal ground state as well as free vector-boson 
excitations whose mass derives from an adjoint Higgs mechanism, representing a tight coupling to the 
thermal ground state. As such, the entire deconfining thermodynamics of 
SU(2)$_{\text{CMB}}$ is influenced by a fixed (Yang-Mills) mass scale $\Lambda_{\text{CMB}}\sim 10^{-4}\,$eV 
\cite{Hofmann2016a}. However, the purely photonic part of SU(2)$_{\text{CMB}}$ is still conformal\footnote{The way how the SU(2)$_{\rm CMB}$ photon gas relates to its thermal ground state 
respects this disconnectness from scale $\Lambda_{\text{CMB}}$ microscopically: Energy 
and momentum quanta are invoked by inert (anti)caloron centers, which themselves are energy and momentum 
free, while a small low-frequency spectral range 
of wave-like excitations only appeals to polarisable electric and magnetic 
dipole densities whose dipole moments and associated volume per dipole moment, again, are determined by inert 
(anti)caloron centers, see \cite{Hofmann2016b}.}, at least on the level of 
free thermal quasiparticle fluctuations which is sufficiently accurate for our purposes. 
Therefore and because of adiabatically slow cosmological expansion the gas of thermal photons solely is governed by (redshift dependent) temperature $T$. As a consequence, a quantity of dimension mass (natural units: $c=\hbar=k_B=1$) 
describing the spectral properties of the thermal photon gas, say, the circular frequency 
$\omega$ of a thermal photon, is expressible as a {\sl redshift independent}, dimensionless multiple $x$ of $T$, 
\eqb
\label{protoomega}
\omega=xT\,.
\eqe

Literature testifies to proposed and actual measurements of the CMB temperature 
$T$ as a function of redshift $z$ by appealing to the thermal Sunyaev-Zel'dovich (tSZ) 
effect ($z\le 1$). This effect represents small negative (low $x$) or positive (high $x$) 
shifts $\Delta I_{\rm tSZ}$ of spectral intensity compared to the CMB black-body spectrum caused by (recoil-free) interaction 
of CMB photons with electrons in high-temperature plasmas ($T_e\sim 10\,$keV) occurring in galaxy clusters, 
see, e.g. \cite{Rephaeli,Luzzi}. Other extractions of $T(z)$ appeal to the excitation of 
molecular rotation bands or atomic hyperfine lines by interaction with the CMB, 
see, e.g. \cite{Muller}. For the tSZ effect one has 
\eqb
\label{sstSZ}
\Delta I_{\rm tSZ}=\frac{T^3_0}{2\pi^2}\frac{x^4\,\e^x}{(\e^x-1)^2}\tau\left(\theta f(x)-v_r+C(x,\theta,v_r)\right)\,,
\eqe 
where $\tau=\sigma_T\int dl\,n_e$ is the optical depth ($\sigma_T$ the Thomson-scattering 
cross section; $\int dl\,n_e$ the electron density, projected through the cluster along the line of 
sight), $\theta\equiv\frac{T_e}{m_e}$ ($m_e$ rest mass of electron), $v_r$ denotes the radial component of the cluster's 
peculiar velocity, and $f(x)\equiv x\coth\frac{x}{2}-4$. Function $C$ describes (small) 
relativistic corrections. $\Delta I_{\rm tSZ}$ can be conceived as the linear term 
in a spectrally local temperature shift 
\eqb
\label{deltaT}
\Delta T_{\rm tSZ}\equiv T_0\tau\left(\theta f(x)-v_r+C(x,\theta,v_r)\right) 
\eqe
when expanding 
spectral intensity about the undistorted black-body spectrum: The factor 
$\frac{T^2_0}{2\pi^2}\frac{x^4\e^x}{(\e^x-1)^2}$ represents the first derivative 
of black-body spectral intensity w.r.t. temperature $T$ at $T=T_0$. 
Notice that $\Delta I_{\rm tSZ}$ depends on the CMB photon circular frequency 
$\omega$ through the dimensionless variable $x$ only whose $z$-independence is a 
consequence of the one-scale (or conformal) nature of 
the undistorted thermal photon gas as discussed above. Furthermore, there are dependences on dimensionless 
quantities, $\theta$ and $v_r$. However, statistically seen, $\theta$ and $v_r$ are not expected to be 
$z$-dependent. $\Delta I_{\rm tSZ}$ thus is 
redshift independent to linear order in $\Delta T_{\rm tSZ}$. Organizing the tSZ effect as an expansion in powers of 
$\Delta T_{\rm tSZ}$, the dependence on 
$\omega$ of the coefficients of higher-than-linear powers in $\Delta T_{\rm tSZ}$ may no 
longer occur solely via $x$. This would violate the exact $z$-independence of $\Delta I_{\rm tSZ}$ at an 
immeasurable level, however. Based on these observations it is clear that the tSZ effect cannot 
be used to extract $T(z)$: The usual prejudice that the circular frequency $\omega$ of a 
thermal CMB {\sl photon} is blueshifted as $\omega=(z+1)\omega_0$ \cite{Luzzi} 
immediately implies that also $T=(z+1)T_0$. This, indeed, is "extracted" from the 
data. Conversely, the theoretical prediction of $T(z)=g(z)T_0$ ($g$ the dimensionless function 
encoded in Fig.\,\ref{Fig-1}), being a consequence of SU(2)$_{\rm CMB}$ energy conservation 
in a FLWR universe 
\cite{Hofmann2015}, immediately implies the according blueshift law 
$\omega=g(z)\omega_0$ for the circular frequency of a CMB photon. Since CMB {\sl photons} in 
contrast to propagating electromagnetic {\sl waves} are {\sl incoherent} 
fluctuations such a blueshift law has no exploitable information content. 
In this context we stress that the observation of redshifts of 
frequencies in atomic emission spectra from astrophysical objects 
are a completely nonthermal affair: These spectra are 
propagated towards the observer by directed, {\sl wave-like} disturbances subject to the 
U(1) Cartan subgroup of an SU(2) gauge group subject to a Yang-Mills 
scale largely disparate from $\Lambda_{\text{CMB}}\sim 10^{-4}\,$eV \cite{Hofmann2016b}. Recall, that the blueshift $\nu=(z+1)\nu_0$ of observed frequency $\nu_0$ is an 
immediate consequence of the emitted wave traveling along a null-geodesic in FLRW 
cosmology. As for the "extraction" of $T(z)$ in terms of the temperature setting the thermally weighted 
population of hyperfine atomic or rotational molecular levels upon radiative coupling of the considered 
species with the CMB, one also assumes that a CMB photon frequency $\nu$, which matches a 
transition frequency, is redshifted according to the conventional 
theory, see, e.g. \cite{Muller}. However, since the frequency dependence of the 
associated column density rests on Boltzmann exponentials of CMB temperature $T$, this prejudice prescribes $T(z)$ in the 
sense discussed above: $T(z)$ necessarily is "extracted" to be conventional. 
As an aside, we point out in Appendix B, Fig.\,\ref{Fig-B1}, that the usual power-law 
parametrisation $T(z)=(1+z)^{1-\beta}T_0$ ($\beta$ fixed) in observational "extractions" of $T(z)$ 
is not satisfactory when confronted with the low-$z$ behaviour of $T(z)$ 
shown in Fig.\,\ref{Fig-1}.   

Because of relation (\ref{T-z}) CMB decoupling sets in at a redshift of $\sim 1800$ which 
is highly disparate from $z\sim 1100$ purported by the $\Lambda$CDM concordance model. 
As discussed in \cite{Hofmann2015}, this suggests that at CMB decoupling the role of non-relativistic matter in $\Lambda$CDM cosmology, 
composed of cold dark and baryonic contributions, is played solely by the baryons. The question then 
arises how the well-tested $\Lambda$CDM model emerges at intermediate redshifts. Based on percolated and unpercolated 
solitons \cite{Wetterich2001} of a Planck-scale axion field (U(1) vortices) Appendix D proposes a possible answer. 
To clarify whether such a scenario can explain the observed rotation curves of spiral galaxies and 
the extensive data on structure formation -- thus seeded by Planckian physics and phase transitions in the early universe -- {\sl much more} work is required.  

The present paper therefore adopts the point of view that for cosmological purposes 
the $\Lambda$CDM model is a useful and accurate approximation to the actual physics 
for $z\lesssim 9$. At the same time, we suspect that for higher values of $z$ this model 
increasingly fails because of relation (\ref{T-z}). Fortunately, our high-$z$ cosmological model, which is conservative 
concerning its (solely baryonic) matter content and the number of massless neutrino flavours 
but invokes SU(2)$_{\text{CMB}}$ to 
describe the CMB itself, can be tested in terms of the most basic low-$z$ cosmological 
parameter: today's value of the cosmic expansion rate $H_0$. This test relies on an 
inverse proportionality between the co-moving sound horizon at baryon freeze-out $r_s$ and $H_0$ which was extracted from 
low-$z$ data (Cepheid calibrated SNe~Ia, new parallax measurements, stronger constraints
on the Hubble flow, and a refined computation of distance to NGC4258 from maser data) under 
no model assumptions other than spatial flatness, SNe~Ia/$r_s$ yielding standard candles/a standard ruler, 
and a smooth expansion history \cite{Bernal2016}. Obviously, this knowledge is important 
because it allows a high-$z$ extraction of $r_s$ to determine $H_0$. 
Note that the value of $H_0$, as obtained by CMB analysis based on $\Lambda$CDM and U(1) photons \cite{Ade2016}, is at $3.4\sigma$ tension with the direct measurement of $H_0$ in \cite{Riess2016}.  

When (re-)computing $r_s$ in the new, high-$z$ SU(2)$_{\text{CMB}}$ based model with $N_\nu=3$ massless neutrinos, $T_0=2.725\,$K, and solely baryonic matter (simply referred to SU(2)$_{\text{CMB}}$ in the following) and in the $\Lambda$CDM model we consider the parameter values of $\eta_{10}, Y_P, N_{\text{eff}}$, and $\Omega_{\text{CDM}}$ of \cite{Ade2016} as 
representative, see Sec.\,\ref{CM}.

This work is organized as follows. In Sec.\,\ref{CM} we introduce SU(2)$_{\text{CMB}}$ and discuss its 
parameter setting. A rough estimate of the decoupling redshift $z_{\text{dec}}$, obtained 
by assuming (i) thermalization (Saha equation) and (ii) instantaneous decoupling, is carried out for 
both SU(2)$_{\rm CMB}$ and $\Lambda$CDM in Sec.\,\ref{SH}. Comparing our 
value for $z_{\text{dec}}$ with the values for $z_*$ and 
$z_{\rm drag}$ in $\Lambda$CDM, we conclude that this 
approximation systematically over-estimates the conventional redshift values 
for decoupling and baryon velocity freeze-out. This led us to perform a 
realistic simulation of the recombination physics in Sec.\,\ref{RTR}
based on the Boltzmann code \texttt{recfast}, the simulation for $\Lambda$CDM serving as a 
check in reproducing the values for $r_s(z_{\rm drag})$ and $r_s(z_{*})$ of 
\cite{Ade2016}. For SU(2)$_{\text{CMB}}$ we find 
that $z_{\rm drag}>z_{*}$. Because the visibility functions in the formal 
solutions of the Boltzmann hierarchy for the temperature perturbations and of the 
Euler equation for baryon velocity are not delta-like, as assumed 
in \cite{Hu1996}, but exhibit considerable widths 
freeze-out redshifts $r_s(z_{\rm lf,drag})$ and $r_s(z_{{\rm lf},*})$ are determined by 
the left flanks of these visibility functions rather than their centers. 
Subsequently, we compute $r_s(z_{{\rm lf},*})$ and $r_s(z_{\rm lf,drag})$. 
With the $r_s-H_0$ relation of \cite{Bernal2016} we deduce from our value of $r_s(z_{\rm lf,drag})$ a good match of 
$H_0$ with the value given in \cite{Riess2016}. In Sec.\,\ref{SO} we summarize our results, briefly discuss a cosmological 
model, which, based on topological solitons of a Planck-scale axion field, 
interpolates our high-$z$ SU(2)$_{\text{CMB}}$ model with low-$z$ $\Lambda$CDM, and sketch a road map for future work. 
Appendix A contains a table documenting the changes in the \texttt{recfast} code 
when adapted to SU(2)$_{\text{CMB}}$. Appendix B addresses 
pecularities in fitting $T(z)$ for SU(2)$_{\text{CMB}}$. Appendix C investigates the 
solution to the Euler equation describing baryon-velocity evolution to 
argue that freeze-out occurs at $z_{\rm lf,drag}$ rather than $z_{\rm drag}$. 
Appendix D provides technical details on a high-$z$ to low-$z$ interpolating cosmological 
model together with a computation of the angular scale $\theta_{*}$ associated with the sound 
horizon at photon freeze-out which is observable in the CMB TT correlation function.   

\section{High-$z$ cosmological model and sound horizon\label{CM}}

Let us introduce our high-$z$ cosmological model SU(2)$_{\text{CMB}}$. As usual, a subscript "0" refers to today's value of the 
associated quantity, we work in super-natural units ($c=\hbar=k_B=1$), and 
we assume a spatially flat FLRW universe such that 
\eqb
\label{Hz}
H(z)=H_0\,\sqrt{\sum_{i}\Omega_i(z)}\,,
\eqe
with $\Omega_i(z=0)\equiv \Omega_{i,0}$ defining the ratio of today's 
energy density $\rho_{i,0}$ of the $i$th separately conserved (and at $z\gg 1$ relevant) cosmic fluid 
to today's critical density $\rho_{c,0}\equiv 3/(8\pi G)\,H_0^2$. Here $G$ denotes Newton's 
constant. Furthermore, we make the convention  
\eqb
\label{H0}
H_0\equiv h 100\,\mbox{km}\,\mbox{s}^{-1}\,\mbox{Mpc}^{-1}\,.
\eqe
We consider $N_\nu=3$ flavours of 
massless neutrinos \cite{Beringer2012} to form a separately conserved cosmic fluid. Because of SU(2)$_{\text{CMB}}$ the neutrino  
temperature $T_\nu$ is determined by the CMB temperature $T$ as in Eq.\,(\ref{TneutTcmbSU2}), and their energy density 
$\rho_\nu$ relates to the energy density of CMB photons $\rho_\gamma$ as 
\eqb
\label{rhonuvsrhogamma}
\rho_\nu(T)=\frac78 \left(\frac{16}{23}\right)^{4/3}N_\nu\rho_\gamma(T)\,.
\eqe 
Because eight instead of two relativistic d.o.f. determine the energy density $\rho_{\text{SU(2)}_{\text{CMB}}}$ at high-$z$ one has 
\eqb
\label{rhoSu2onT} 
\rho_{\text{SU(2)}_{\text{CMB}}}(T)=4\,\rho_\gamma(T)\,.
\eqe
Writing $\rho_{\text{SU(2)}_{\text{CMB}}}$ as a function of 
$z$, Eq.\,(\ref{T-z}) implies 
\eqb
\label{rhoSu2onz} 
\rho_{\text{SU(2)}_{\text{CMB}}}(z)
=A\,\rho_{\gamma,0} (z+1)^4\,,
\eqe
where 
\eqb
\label{Adef}
A\equiv (8/2) \cdot (0.63)^4 \,.
\eqe 
Considering Eq.\,(\ref{Hz}), Eqs.\,(\ref{rhoSu2onz}) and (\ref{rhonuvsrhogamma}) are recast as 
\eqb
\label{rhoSu2onzOmega} 
\rho_{\text{SU(2)}_{\text{CMB}}}(z)
=A\,H_0^2\,\Omega_{\gamma,0}(z+1)^4\frac{3}{8\pi G}\,,
\eqe
and 
\eqb
\label{rhonuvsrhogammaOmega} 
\rho_\nu(z)
=\frac{7}{8}\frac{A}{4}\left(\frac{16}{23}\right)^{4/3}\,N_{\nu}\,H_0^2\,\Omega_{\gamma,0}(z+1)^4\frac{3}{8\pi G}\,.
\eqe 
Setting $T_0=2.725\,$K \cite{Fixsen1996}, one has $\Omega_{\gamma,0}=2.46796\times 10^{-5}\,h^{-2}$. 

Non-relativistic, cosmological matter is assumed to be purely baryonic in SU(2)$_{\text{CMB}}$. 
Its energy density today, $\rho_{b,0}$, relates to $\rho_{\gamma,0}$ as
\eqb
\label{rhobrhogamma} 
\rho_{b,0}=\frac43 R_0\,\rho_{\gamma,0}\,
\eqe  
such that 
\eqb
\label{Rofeta} 
R_0\equiv 111.019\,\eta_{10}\,,
\eqe
where 
\eqb
\label{Rdef}
R\equiv \frac{3}{4}\frac{\rho_b}{\rho_\gamma}\,.
\eqe
From Big-Bang Nucleosynthesis (BBN) $\eta_{10}$ is constrained to $4.931\le\eta_{10}\le 7.123$, see Fig.\,29 in \cite{Ade2014b}. The number $\eta_{10}$ parametrises today's baryon-to-photon number-density ratio $n_{b,0}/n_{\gamma,0}$ as 
\eqb
\label{btoph}
n_{b,0}/n_{\gamma,0}\equiv \eta_{10}\times 10^{-10}\,.
\eqe 
Note that the central value in 
\eqb
\label{eta10}
\eta_{10}=6.08232\pm 0.06296\,, 
\eqe 
as computed from the value $\Omega_{b,0}=(0.02222\pm 0.00023)\,h^{-2}$ obtained 
by the Planck collaboration \cite{Ade2016}, is also central to the 
above BBN range. According to \cite{Ade2016} this 
value of $\eta_{10}$ implies a $\mbox{}^4$He mass fraction $Y_P$ of
\eqb
\label{YP}
Y_P=0.252\pm 0.041\,.
\eqe
Note that due to non-conformal $T$-$z$ scaling in SU(2)$_{\text{CMB}}$ there is a low-$z$ dependence of 
$n_{b}/n_{\gamma}$ -- in contrast to the conventional case of U(1) photons. Also, because of 
Eq.\,(\ref{T-z}) the high-$z$ expression for quantity $R$ reads
\eqb
\label{RofetaZ} 
R(z)\equiv 111.019\,\frac{\eta_{10}}{(0.63)^4(z+1)}\,. 
\eqe 
Taking the central value for $\eta_{10}$ from Eq.\,(\ref{eta10}), appealing to 
$T_{\text{dec}}\sim 3000\,$K, and considering Eq.\,(\ref{T-z}), we roughly estimate 
the redshift $z_{\text{dec}}$ 
for the end  of hydrogen recombination as 
$z_{\text{dec}}\sim 3000/(0.63\cdot 2.725)-1\sim 1775$. Therefore, we conclude from 
Eq.\,(\ref{Rofeta}) that in SU(2)$_{\text{CMB}}$ $R(z)>1$ for $z$ ranging from $z=0$ to well beyond recombination: 
$R(z)>1$ for $z<4568$. This is in contrast 
to the conventional $\Lambda$CDM model where CMB decoupling occurs at 
$z_{\text{dec}}\sim 1100$ and where $R$ is given as 
\eqb
\label{RofetaLambda} 
R(z)\equiv 111.019\,\frac{\eta_{10}}{z+1}\, 
\eqe  
such that $R(z)<1$ for $z>675$. As usual we have 
\eqb
\label{}
\frac{\rho_{b}(z)}{\rho_{c,0}}\equiv \Omega_{b,0}(z+1)^3\,,
\eqe 
where \cite{Ade2016}
\eqb
\label{Omegabvseta10}
\Omega_{b,0}\equiv 0.00365321\,\eta_{10}\,h^{-2}\,.
\eqe
Therefore, in SU(2)$_{\text{CMB}}$ the high-$z$ Hubble parameter $H$ reads 
\begin{align}
H(z)=&H_0\,\left[\Omega_{b,0}\,(z + 1)^3 +\right.\nonumber\\
&\left. A\left((1+\frac{7}{32}\left(\frac{16}{23}\right)^{4/3}\,N_{\nu}\right)
\Omega_{\gamma,0}(z + 1)^4\right]^{1/2}\,,\nonumber\\ 
\label{Hzexp}
\end{align}
where $\Omega_{b,0}$ and its errors derive from $\eta_{10}$ as  
quoted in Eq.\,(\ref{eta10}). We also consider the high-$z$ $\Lambda$CDM model 
\begin{align}
\label{Hzexpconv}
H(z)=&H_0\,\left[\left(\Omega_{b,0}+\Omega_{\text{CDM}}\right)\,(z + 1)^3 +\right.\nonumber\\ 
&\left.\left(1+\frac{7}{8}\left(\frac{4}{11}\right)^{4/3}\,N_{\text{eff}}\right)
\Omega_{\gamma,0}(z + 1)^4\right]^{1/2}\,,\nonumber\\ 
\end{align}
where according to \cite{Ade2016} we have 
\eqb
\label{Neffetc}
\Omega_{\text{CDM}}=(0.1197\pm 0.0022)\,h^{-2}\,,\ \ \ N_{\text{eff}}=3.15\pm 0.23\,. 
\eqe 
Note that in both cases, Eqs.\,(\ref{Hzexp}) and (\ref{Hzexpconv}), 
the high-$z$ expressions for $H(z)$  
are independent of $h$. An overview of the differences between high-$z$ $\Lambda$CDM and 
SU$(2)_{\text{CMB}}$ is presented in Tab.\,\ref{cosmomodels}. 

\begin{table}
	\centering
	\caption{Cosmological high-$z$ models: $\Lambda$CDM versus SU$(2)_{\text{CMB}}$.\label{cosmomodels}}
	\begin{tabular}{CCC}
	& & \\
		\toprule
		& \Lambda\text{CDM} & \text{SU(2)}_{\text{CMB}} \\
		\midrule
		\frac{T}{T_0} & z+1 & 0.63\,(z+1) \\
		\frac{T_\nu}{T} & \left(\frac{4}{11}\right)^{1/3}  & \left(\frac{16}{23}\right)^{1/3} \\[0.35cm]
		\Omega_{\text{CDM}} & \Omega_{\text{CDM}} & 0 \\
		N_\nu & N_{\text{eff}} & 3 \\
		\bottomrule
	\end{tabular}
\end{table}

The co-moving sound horizon $r_s(z)$, as emergent within the baryon-electron-photon plasma, is defined as
\eqb
\label{shdef}
r_s(z)=\int_0^{\eta(z)} d\eta^\prime\,c_s(\eta^\prime)=\int_z^\infty dz^\prime\,\frac{c_s(z^\prime)}{H(z^\prime)}\,,
\eqe
where $\eta$ is conformal time ($\mathrm{d}\eta\equiv \mathrm{d}t/a$), and $c_s$ denotes the sound velocity, given 
as 
\eqb
\label{soundvel}
c_s\equiv\frac{1}{\sqrt{3(1+R)}}\,.
\eqe
In Eq.\,(\ref{soundvel}) $R$ either needs to be taken from Eq.\,(\ref{RofetaZ}) (SU(2)$_{\text{CMB}}$) or from 
Eq.\,(\ref{RofetaLambda}) ($\Lambda$CDM). 

Finally, we would like to explain how we perform error estimates for $r_s(z)$. For example, in 
SU(2)$_{\text{CMB}}$ error-prone input parameters are $\eta_{10}$ and $Y_P$. For those we 
generate pairs of Gaussian distributed random values. For each pair we compute 
$r_s(z)$ and fit a Gaussian to the ensuing histogram in order to extract the 1-$\sigma$ error range for  
$r_s(z)$. In doing this, $z$ needs to satisfy a condition, specified, e.g. by either  Eqs.\,(\ref{cheGammH}), 
(\ref{optdepthThomson}), or (\ref{zdragdef}), to determine its value $z_{\text{dec}}$, $z_*$, and $z_{\text{drag}}$, respectively. For $\Lambda$CDM the set $\{\eta_{10}, Y_P\}$ is enhanced by the elements $\Omega_{\text{CDM}}$ 
and $N_{\text{eff}}$. For an overview of the values of the cosmological parameters see Tab.\,\ref{TabcsomPara}.

\begin{table*}
	\centering
	\caption{Cosmological parameter values employed in the computations and their sources.\label{TabcsomPara}}
	\begin{tabular}{lcl}
	   & & \\
		\toprule
		parameter & value  & source  \\
		\midrule
		$H_0$ (SU$(2)_{\text{CMB}}$) & $(73.24\pm 1.74)$\,km\,s$^{-1}$\,Mpc$^{-1}$  & \cite{Riess2016}  \\
		$H_0$ ($\Lambda$CDM) & $(67.31\pm 0.96)$\,km\,s$^{-1}$\,Mpc$^{-1}$  & TT+lowP, \cite{Ade2016}   \\
		$T_0$ & 2.725\, K &  \cite{Fixsen1996}\\
		$\Omega_{\gamma,0} h^2$ & $2.46796\times 10^{-5}$.  & based on $T_0=2.725\,$K \\
		$\Omega_{b,0} h^2$ & $0.02222 \pm 0.99923$ & TT+lowP, \cite{Ade2016} \\
		$\Omega_{\text{CDM},0} h^2$ & $0.1197\pm 0.0022$& TT+lowP, \cite{Ade2016} \\
		$\eta_{10}$ & $6.08232\pm 0.06296$ & based on $\Omega_{\gamma,0}h^2$, TT+lowP, \cite{Ade2016}  \\
		$Y_P$ & $0.252\pm 0.041$ & TT, \cite{Ade2016}\\ 
		$N_{\text{eff}}$ & $3.15\pm 0.23$ & abstract, \cite{Ade2016} \\ 
		\bottomrule
	\end{tabular}
\end{table*}

\section{Saha equation and instantaneous CMB decoupling/radiation drag\label{SH}}

Before we turn to a detailed analysis of recombination physics in Sec.\,\ref{RTR}, 
let us now perform a rough estimate for a single redshift $z_{\text{dec}}$ associated with 
CMB decoupling physics/radiation drag (baryon velocity freeze-out). In the present section, we base our 
estimate on two assumptions: (i) thermalization (Saha equation) 
and (ii) coincidence of decoupling and radiation drag, both of vanishing duration. 

Appealing to the results of 
\cite{Bernal2016} on the low-$z$ inverse proportionality between the sound horizon $r_s$,  
seen in today's baryonic matter correlation, and $H_0$, our here-determined 
central value of $H_0$ for $\Lambda$CDM over-estimates the 
result 
\eqb
\label{Planck2015}
H_0=(67.31\pm 0.96)\,\mbox{km\,s}^{-1}\,\mbox{Mpc}^{-1}
\eqe 
of \cite{Ade2016}. Also, our estimate of $H_0$ for SU(2)$_{\text{CMB}}$ 
is higher than the directly measured value 
\eqb
\label{adam}
H_0=(73.24\pm 1.74)\,\mbox{km\,s}^{-1}\,\mbox{Mpc}^{-1}
\eqe  
of \cite{Riess2016}. This motivates our analysis of Sec.\,\ref{RTR}. 

In the present section, the value of $z_{\text{dec}}$ is determined from condition  
\eqb
\label{conddecsimp}
H(z_{\text{dec}})=\Gamma(z_{\text{dec}})\,.
\eqe
In Eq.\,(\ref{conddecsimp}) the rate $\Gamma$ for scattering of eV-photons off free, non-relativistic electrons reads 
\eqb
\label{defGamma}
\Gamma=\sigma_T n_e^b \chi_e\,,
\eqe 
where $\sigma_T\equiv 6.65\times 10^{-25}\,$cm$^2$ denotes the 
Thomson cross section for electron-photon scattering, and $n_e^b$ is the electron 
density just before the onset of hydrogen recombination which
is given as  
\begin{align}
n_e^b(z) &\equiv(1-Y_P)n_b(z) \nonumber\\
&=410.48\cdot10^{-10}\,\eta_{10}(1-Y_P)(z+1)^3\,\mbox{cm}^{-3} \label{neb}\,. 
\end{align}
Moreover, $\chi_e$ refers to the ionization fraction during the recombination epoch, 
\eqb
\label{iondef}
\chi_e(z)\equiv \frac{n_e(z)}{n_e^b(z)}\,,
\eqe
$n_e$ being the actual electron density, evolving non-trivially during 
recombination, see Sec.\,\ref{RTR}. 
In our present treatment we set $z=z_{\text{dec}}$ in Eqs.\,(\ref{neb}) and (\ref{iondef}). 
We also use the Saha equation, which assumes thermal equilibrium 
between electrons, photons, and ions,\footnote{Thomson scattering off neutral hydrogen and Helium atoms can safely be neglected \cite{Mukhanov2005}.}  
\eqb
\label{Saha}
\frac{\chi^2_e}{1-\chi_e}=
\frac{1}{n_e^b}\left(\frac{T_{\text{dec}} m_e}{2\pi}\right)^{3/2}\,\exp\left(-\frac{B_H}{T_{\text{dec}}}\right)\equiv S\,,
\eqe
to estimate $\chi^2_e$ at $z_{\text{dec}}$. In Eq.\,(\ref{Saha}) the following values are set for the quantities $m_e, B_H$:
\eqb
\label{valesaha}
m_e=510998.94\,\mbox{eV}\,,\ \ \ B_H = 13.6\,\mbox{eV}\,.
\eqe
Depending on whether the cosmological model of Eq.\,(\ref{Hzexp}) (SU(2)$_{\text{CMB}}$) or Eq.\,(\ref{Hzexpconv}) ($\Lambda$CDM) is considered, we set in Eq.\,(\ref{Saha})
$T_{\text{dec}}=0.63(z_{\text{dec}}+1)T_0$ or $T_{\text{dec}}=(z_{\text{dec}}+1)T_0$, respectively. Solving Eq.\,(\ref{Saha}) for $\chi_e$, we have
\eqb
\label{chesaha}
\chi_e=\frac12\left[-S + S^{1/2}(4+S)^{1/2}\right]\sim S^{1/2}\ \ \ \ (S\ll 1)\,.
\eqe 
On the other hand, solving Eq.\,(\ref{conddecsimp}) for $\chi_e$ yields
\eqb
\label{cheGammH}
\chi_e(z_{\text{dec}})=\frac{1}{\sigma_T n_e^b(z_{\text{dec}})}H(z_{\text{dec}})\,,
\eqe
where either the expression in Eq.\,(\ref{Hzexp}) (SU(2)$_{\text{CMB}}$) 
or in Eq.\,(\ref{Hzexpconv}) ($\Lambda$CDM) is substituted for $H$. Equating the right-hand sides of Eq.\,(\ref{chesaha}) 
and Eq.\,(\ref{cheGammH}) as foreseen by Eq.\,(\ref{conddecsimp}), 
we derive approximate values for $z_{\text{dec}}$ and their errors from 
$\eta_{10}, Y_P, \Omega_{\text{CDM}}$, and $N_{\text{eff}}$ as quoted in Eqs.\,(\ref{eta10}), (\ref{YP}), and (\ref{Neffetc}), respectively, see also Tab.\,\ref{TabcsomPara}. We obtain
\begin{align}
\label{su2lambdacdm}
z_{\text{dec}}&=1760.14\pm 1.85\ \ \ \ (\mbox{SU(2)}_{\text{CMB}})\,,\nonumber\\ 
z_{\text{dec}}&=1132.78\pm 1.27\ \ \ \ (\Lambda\mbox{CDM})\,.
\end{align}

\begin{figure}
\centering
\includegraphics[width=\columnwidth]{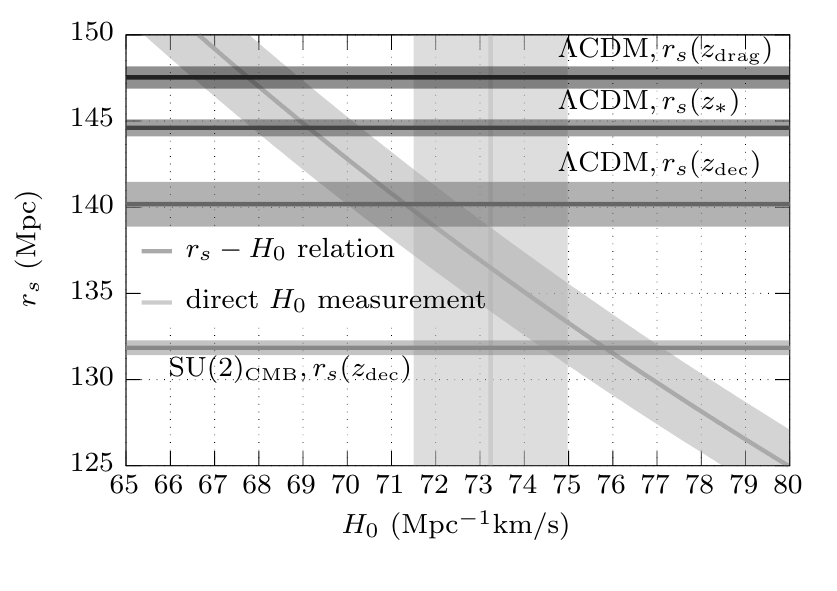}
\caption{\protect{\label{Fig-2}}Instantaneous-decoupling predictions of the sound horizon 
$r_s$ including the 1-$\sigma$ error range in high-$z$ $\Lambda$CDM (third horizontal band) and SU(2)$_{\text{CMB}}$ (fourth horizontal band) together 
with the low-$z$ $r_s$-$H_0$ relation of \protect\cite{Bernal2016} (curved band) and the direct measurement of 
$H_0$ as reported in \protect\cite{Riess2016} (vertical band). For 
completeness we also quote $r_s(z_{\text{drag}})$ (first horizontal band) and $r_s(z_{*})$ (second horizontal band)
in $\Lambda$CDM, for definitions see Sec.\,\ref{RTR}.}     
\end{figure}

Appealing to Eq.\,(\ref{shdef}), we arrive at 
\begin{align}
\label{su2lambdacdmrs}
r_s(z_{\text{dec}})&=(131.85\pm 0.43)\,\mbox{Mpc}\ \ \ \ (\mbox{SU(2)}_{\text{CMB}})\,,\nonumber\\ 
r_s(z_{\text{dec}})&=(140.18\pm 1.30)\,\mbox{Mpc}\ \ \ \ (\Lambda\mbox{CDM})\,,
\end{align}
see Fig.\,\ref{Fig-2}. Amusingly, the intersections of the bands $r_s(z_{\text{dec}})$ in 
SU(2)$_{\text{CMB}}$ and $\Lambda$CDM with the $r_s-H_0$ band of \cite{Bernal2016} have a non-vanishing intersection 
with the 1-$\sigma$ range of $H_0$ measured in \cite{Riess2016}. However, we observe that $r_s(z_{\text{dec}})$ is considerably under-estimated compared to $r_s(z_{\text{drag}})$ in $\Lambda$CDM. Therefore, we suspect that $r_s(z_{\text{dec}})$ is also under-estimated in SU(2)$_{\text{CMB}}$ compared to its true value at baryon freeze-out.  
Indeed, since $\chi_e(z_{\text{dec}})=0.003$ (SU(2)$_{\text{CMB}}$) 
and $\chi_e(z_{\text{dec}})=0.010$ ($\Lambda$CDM) we are 
left with considerable doubt on whether our present treatment yields reliable 
results.

\section{Realistic treatment of recombination\label{RTR}}

\begin{figure}
\centering
\includegraphics[width=\columnwidth]{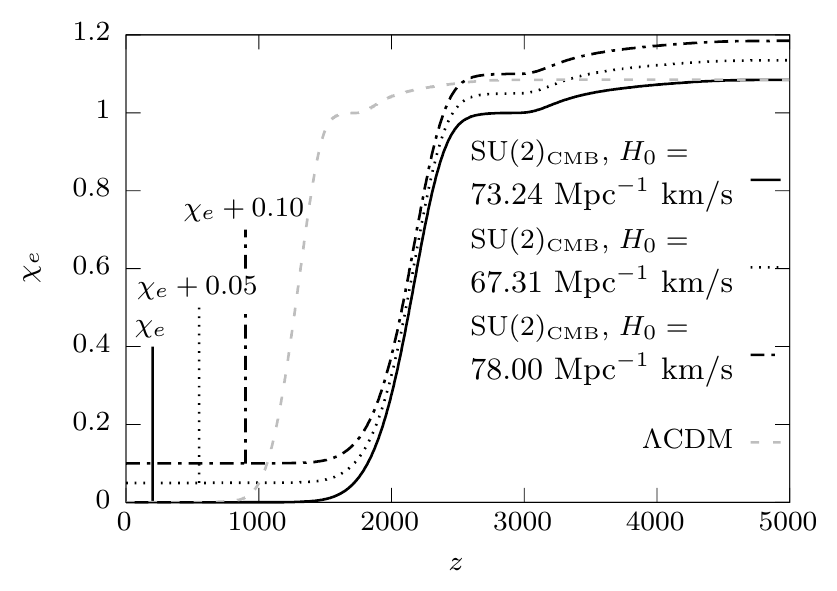}
\caption{Histories for the ionization fraction $\chi_e(z)$ in 
SU(2)$_{\text{CMB}}$ (black with artificially introduced  
vertical offsets to distinguish curves for different values of $H_0$)  
and $\Lambda$CDM (gray, dashed) subject to the parameter values defined in Tab.\,\ref{TabcsomPara}, see also Sec.\,\ref{CM}. 
Regions, for which $\chi_e(z)>1$, associate with incomplete Helium recombination. 
\protect{\label{Fig-3}}}      
\end{figure}

Here we would like to subject recombination physics to realistic histories of the ionization fraction
$\chi_e(z)$. We appeal to the publically available Boltzmann code \texttt{recfast} \cite{Seager1999,Seager2000,Wong2008,Scott2009} which also was used in 
\cite{Ade2014b}. When computing $\chi_e(z)$ in SU(2)$_{\text{CMB}}$ the following code adjustments need to 
be performed: re-set fnu from fnu=$21/8\times(4/11)^{4/3}$ 
to fnu=$21/8\times(23/16)^{4/3}$ ($N_{\text{eff}}=3=N_\nu$ by default) 
and re-define ranges in $z$ for treatments by Saha, Peebles, or Boltzmann equation through divisions  
by 0.63. Note that for a fixed value of 
$\Omega_{b,0}$ (and $\Omega_{\text{CDM}}=0$) the value $H_0$ can be varied in association with 
value of $\Omega_{\Lambda}$ such that the curvature term in 
$H(z)$ is nil. For an exposition of important changes when going from $\Lambda$CDM to SU(2)$_{\text{CMB}}$, 
see Tab.\,\ref{codetable} in Appendix A. Our results for $\chi_e(z)$ do not depend on 
$H_0$ within a reasonable range, see Fig.\,\ref{Fig-3}.

The end of recombination at $z_*$ is usually defined by the optical 
depth $\tau(z_*)$ to Thomson scattering from $z=0$ to $z_*$ being equal to unity 
\cite{Ade2014b}. That is, 
\eqb
\label{optdepthThomson}
\tau(z_*)=\sigma_T\int_0^{z_*}dz\,\frac{\chi_e(z)n_e^b(z)}{(z+1)H(z)}=1\,,
\eqe
where $n_e^b(z)$ and $\chi_e(z)$ are defined in Eq.\,(\ref{neb}) and Eq.\,(\ref{iondef}), 
respectively, and $H(z)$ either is given by Eq.\,(\ref{Hzexp}) (SU(2)$_{\text{CMB}}$) 
or by Eq.\,(\ref{Hzexpconv}) ($\Lambda$CDM). Due to a considerable width 
of a bump-like weight function $D_*$ (here referred to as visibility function) in the formal solution of the according 
Boltzmann hierarchy this criterion\footnote{If $D_*$ had a vanishing width then criterion (\ref{optdepthThomson}) would be applicable.} is revised in Appendix C. As a consequence, the position $z_{{\rm lf},*}$ of the 
left flank of $D_*$ is more realistic for photon decoupling.    

We also consider the standard definition for 
the end of the radiation-drag 
epoch \cite{Hu1996}, relying on the radiation-drag depth $\tau_{\rm drag}(z)$, defined as 
\eqb
\label{dragdepth}
\tau_{\rm drag}(z)=\sigma_T\int_0^{z}dz^\prime\,\frac{\chi_e(z^\prime)n_e^b(z^\prime)}{(z^\prime+1)H(z^\prime)R(z^\prime)}\,.
\eqe 
In \cite{Hu1996} the condition for freeze-out of baryonic velocity at $z_{\text{drag}}$ 
is pronounced to be 
\eqb
\label{zdragdef}
\tau_{\rm drag}(z_{\rm drag})=1\,.
\eqe 
On the basis of the according Euler equation for the baryon-photon fluid we 
show in Appendix C, however, that the baryon velocity $v_b$ is not yet frozen out at 
$z_{\rm drag}$. Roughly speaking, the solution $a v_b$ of the Euler 
equation amounts to an $z^\prime$-integral of $D_{\rm drag}$ (drag visibility function) resembling a bump-like function of 
{\sl finite} width. Notice that $D_{\rm drag}$ was characterised 
as a delta function in \cite{Hu1996}. Freeze-out, that is, $z$-independence of 
this integral, $z$ being the lower integration limit, occurs if $z$ is placed 
sufficiently far to the left of the position $z_{\rm max,drag}$ of the maximum. 
A characteristic point $z_{\rm lf, drag}$ setting a realistic cutoff for the 
$z^\prime$-integration is the position of the left flank defined through the position of 
the maximum of the $z^\prime$-derivative of 
$D_{\rm drag}$. Interestingly, $z_{\rm drag}$ of Eq.\,(\ref{zdragdef}) and 
$z_{\rm max, drag}$ practically coincide. This would support the definition of baryon 
velocity freeze-out in Eq.\,(\ref{zdragdef}) as in \cite{Hu1996}) were it not for the finite width of 
$D_{\rm drag}$. It is this finite width, however, which implies a 
substantial decrease from $z_{\rm drag}$ to $z_{\rm lf,drag}$ in the redshift for baryon-velocity decoupling, 
see Appendix C for the technical argument and further discussions. 

Using the parameter values of Sec.\,\ref{CM} and appealing to Eqs.\,(\ref{optdepthThomson}), (\ref{shdef}), and Fig.\,C2 in 
Appendix C, we obtain
\begin{align}
\label{su2lambdacdm*}
z_{*}&= 1693.55\pm 6.98\ \ \ \ (\mbox{SU(2)}_{\text{CMB}})\,,\nonumber\\ 
z_{{\rm lf},*}&=1554.89\pm 5.18  \ \ \ \ (\mbox{SU(2)}_{\text{CMB}})\,,\nonumber\\ 
z_{*}&= 1090.09\pm 0.42\ \ \ \ (\Lambda\mbox{CDM}), \nonumber\\ 
z_{{\rm lf},*}&=\phantom{1}987.98\pm 3.28 \ \ \ \ (\Lambda\mbox{CDM})\,,
\end{align}
and  
\begin{align}
\label{su2lambdacdmrs*}
r_s(z_{*})&=(135.35\pm 0.52)\,\mbox{Mpc}\ \ \ \ (\mbox{SU(2)}_{\text{CMB}})\,,\nonumber\\ 
r_s(z_{{\rm lf},*})&=(143.34\pm 0.42)\,\mbox{Mpc}\ \ \ \ (\mbox{SU(2)}_{\text{CMB}})\,,\nonumber\\ 
r_s(z_{*})&=(144.61\pm 0.49)\,\mbox{Mpc}\ \ \ \ (\Lambda\mbox{CDM}),\nonumber\\ 
r_s(z_{{\rm lf},*})&=(153.05\pm 3.35)\,\mbox{Mpc}\ \ \ \ (\Lambda\mbox{CDM})\,.
\end{align}
On the other hand, Eqs.\,(\ref{dragdepth}), (\ref{shdef}), and Fig.\,C1 in Appendix C yield
\begin{align}
\label{su2lambdacdmdrag}
z_{\rm drag}&=1812.66\pm 7.01\ \ \ \ (\mbox{SU(2)}_{\text{CMB}})\,,\nonumber\\ 
z_{\rm lf,drag}&=1659.30\pm 5.48 \ \ \ \ (\mbox{SU(2)}_{\text{CMB}})\,,\nonumber\\ 
z_{\rm drag}&=1059.57\pm 0.46 \ \ \ \ (\Lambda\mbox{CDM})\,,\nonumber\\ 
z_{\rm lf,drag}&=\phantom{1}973.12\pm 3.06 \ \ \ \ (\Lambda\mbox{CDM})\,,
\end{align}
and  
\begin{align}
r_s(z_{\rm drag})&=(129.22\pm 0.52)\,\mbox{Mpc}\ \ \ \ (\mbox{SU(2)}_{\text{CMB}})\,,\nonumber\\ 
r_s(z_{\rm lf,drag})&=(137.19\pm 0.45)\,\mbox{Mpc}\ \ \ \ (\mbox{SU(2)}_{\text{CMB}})\,,\nonumber\\ 
r_s(z_{\rm drag})&=(147.33\pm 0.49)\,\mbox{Mpc}\ \ \ \ (\Lambda\mbox{CDM})\,,\nonumber\\ 
r_s(z_{\rm lf,drag})&=(154.57\pm 3.33)\,\mbox{Mpc}\ \ \ \ (\Lambda\mbox{CDM})\,.
\end{align}
\label{su2lambdacdmrsdrag}
 
\begin{figure}
\centering
\includegraphics[width=\columnwidth]{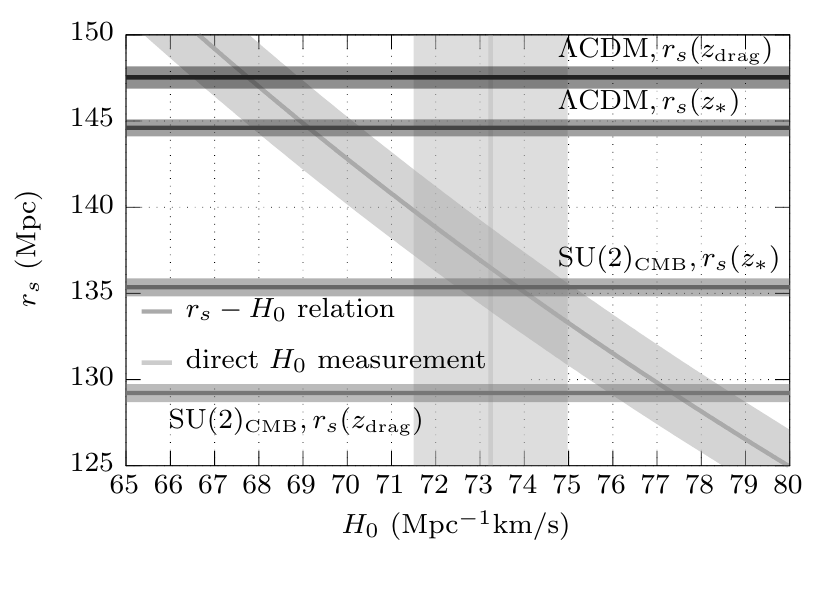}
\caption{\protect{\label{Fig-4}}Predictions of the sound horizon 
$r_s$ including the 1-$\sigma$ error at $z_*$ (second and third horizontal bands) and $z_{\rm drag}$ 
(first and fourth horizontal bands) in high-$z$ $\Lambda$CDM and SU(2)$_{\rm CMB}$, respectively. 
Also shown are the low-$z$ $r_s$-$H_0$ relation of \protect\cite{Bernal2016} (curved band) and the direct measurement of 
$H_0$ (vertical band) as reported in \protect\cite{Riess2016}.}      
\end{figure}

\begin{figure}
\centering
\includegraphics[width=\columnwidth]{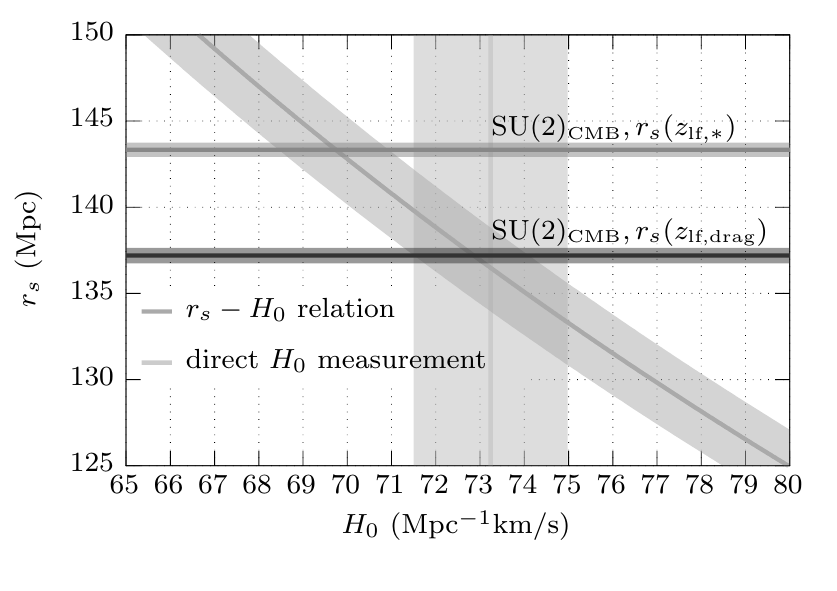}
\caption{\protect{\label{Fig-5}}Prediction of the sound horizon 
$r_s$ including the 1-$\sigma$ error at  $z_{\rm lf,drag}$ (second horizontal band) 
and $z_{{\rm lf},*}$ (first horizontal band) in high-$z$ SU(2)$_{\text{CMB}}$ together 
with the low-$z$ $r_s$-$H_0$ relation of \protect\cite{Bernal2016} (curved band) and the direct measurement of 
$H_0$ (vertical band) as reported in \protect\cite{Riess2016}.}      
\end{figure}

For SU(2)$_{\rm CMB}$ we have (central values of $z_{{\rm lf},*}$ and $z_{\rm lf,drag}$ only)
\eqb
\label{chireal}
\chi_e(z_{{\rm lf},*})\sim 0.013\,,\ \ \ \ \ \ \ \ \ \ \chi_e(z_{\rm lf,drag})\sim 0.032\,.
\eqe
Fig.\,\ref{Fig-4} indicates that, while the intersection of the SU(2)$_{\rm CMB}$-band for 
$r_s(z_{\rm drag})$ with the $r_s-H_0$ band of \cite{Bernal2016} 
is off the 1-$\sigma$ range of the directly measured value of $H_0$ 
\cite{Riess2016}, the intersection of the SU(2)$_{\rm CMB}$-band 
$r_s(z_{*})$ is well contained within this 1-$\sigma$ range. According to our 
discussion in Appendix C, however, none of these statements can be 
considered physical due to imprecise freeze-out conditions.  
Rather, we argue that, due to the finite widths of $D_{\rm drag}$ and 
$D_{*}$, baryon-velocity and photon decoupling occur at the lower values $z_{{\rm lf},*}$ and 
$z_{\rm lf,drag}$ indicated in Eqs.\,(\ref{su2lambdacdmrs*}) and (\ref{su2lambdacdmdrag}), 
respectively. Fig.\,\ref{Fig-5} indicates that the intersection of the SU(2)$_{\rm CMB}$-band for 
$r_s(z_{\rm lf,drag})$ with the $r_s-H_0$ band of \cite{Bernal2016}, indeed, has an impressively large overlap 
with the 1-$\sigma$ range of the $r_s-H_0$ band determined in \cite{Bernal2016}. 
In comparing Figs.\,\ref{Fig-4} and \ref{Fig-5} or by inspecting Eq.\,(\ref{su2lambdacdmrs*}), notice also that $r_s$ at photon-decoupling redshift $z_{{\rm lf},*}$ is about 8\,Mpc larger than 
$r_s$ at $z_{*}$. 

\section{Summary and Outlook\label{SO}}

In the present work we have investigated whether a 3.4-$\sigma$ discrepancy \cite{Bernal2016} in 
the value of the present Hubble parameter $H_0$ can be resolved under minimal assumptions concerning 
the high-$z$ matter sector. This discrepancy relates to the value of $H_0$ as 
extracted by the Planck collaboration under an assumed all-$z$ validity of the $\Lambda$CDM concordance model 
\cite{Ade2016} and the value directly measured in \cite{Riess2016}. As suggested by our results, 
a new, high-$z$ cosmology, which assumes SU(2)$_{\rm CMB}$ thermodynamics \cite{Hofmann2016a}, 
solely baryonic matter, and $N_\nu=3$ species of massless neutrinos, is a candidate. 
Our present analysis was enabled by a model-independent extraction 
of the $r_s$-$H_0$ relation ($r_s$ the co-moving sound horizon at baryon-velocity freeze-out, observable in today's 
matter correlation function) which is based on low-$z$ observation \cite{Bernal2016}. 

Interestingly, in the new model the redshift $z_*$, traditionally thought to set the end of 
hydrogen recombination, is preceded by the redshift $z_{\rm drag}$ proposed in \cite{Hu1996} 
to set the termination of the Compton drag effect. However, due to considerable widths of 
the visibility functions in the formal solutions of the Boltzmann hierarchy for the 
temperature perturbation and the Euler equation for baryon velocity we propose lower 
values $z_{{\rm lf},*}$ and $z_{\rm lf,drag}$ than $z_*$ and $z_{\rm drag}$. 
Using $z_{\rm lf,drag}$ in the computation of $r_s$ 
and the $r_s$-$H_0$ relation of Ref.\,\cite{Bernal2016}, we obtain good agreement with the directly measured 
value of $H_0$ \cite{Riess2016}, see Fig.\,\ref{Fig-5}.

As discussed in \cite{Bernal2016}, the errors of such a direct measurement 
of $H_0$ will shrink substantially in the near future. Thus the here-proposed 
high-$z$ cosmology will soon undergo increasingly stringent tests. 
It remains to be investigated what the influence of this model on higher acoustic harmonics and the associated 
damping physics is. In order to decide this, a cosmological 
model, which interpolates high-$z$ SU(2)$_{\text{CMB}}$ with 
low-$z$ $\Lambda$CDM, is required. Based on percolated and unpercolated topological solitons of a Planck-scale axion 
field \cite{Frieman1995,Neubert} we make a proposal for rough features of such a 
model in Appendix D. A first test -- predicting the angular scale $\theta_*$ 
of the sound horizon at photon decoupling -- yields consistency. This encourages the future computation of 
high-$l$ CMB power spectra (anisotropies and polarization) to provide further tests of SU(2)$_{\text{CMB}}$. 
However, sophisticated routines to compute the CMB power spectra 
like \texttt{CMBfast} or \texttt{CAMB} owe their 
efficiency to a Green's function approach which, in turn, draws on the simplicity of the equations of state of the cosmological fluids in $\Lambda$CDM. Since SU(2)$_{\text{CMB}}$ 
is subject to a complicated equation of state at low z it is not clear whether such a Green's function approach 
is feasible at all. Rather, we would expect that an old-fashioned slice-to-slice evolution is required. 
Thus a quick-shot run of \texttt{CMBfast} or \texttt{CAMB} under questionable approximations is not trustworthy and hence not conclusive. Therefore, together with the pressing importance of developing an 
observationally sound interpolating cosmological model along the lines sketched above, a substantial revision of 
the simulational approach to CMB power spectra is in order. Finally, we would also like to mention here that SU(2)$_{\text{CMB}}$ 
has a potential to successfully address the observed large-angle 
anomalies of the CMB. Namely, as outlined in 
\cite{Hofmann2013} and detailed in \cite{Hofmann2016a}, radiative effects in SU(2)$_{\text{CMB}}$ (transverse and longitudinal contributions to the 
photon polarization tensor $\Pi_{\mu\nu}$), which are important for redshifts $0\le z\le 2$, induce a systematic
departure from statistical isotropy. This is reflected by the build-up 
of a cosmologically local temperature depression (due to the transverse part of $\Pi_{\mu\nu}$), defining a gradient to its 
slope. The associated mild breaking of isotropy in the CMB temperature map would influence the low lying 
CMB multipoles and create intergalactic magnetic fields (due to the longitudinal 
part of $\Pi_{\mu\nu}$).

\section*{Acknowledgements}

We would like to thank Jose Luis Bernal and Adam Riess for providing us with their data files on the low-$z$ 
$r_s-H_0$ relation and the direct measurement of $H_0$, and we would like to 
acknowledge the induction of a rewarding revision process by a very constructive and competent Reviewer. 
Moreover, we thank the Centre National de la Recherche 
Scientifique (CNRS) for the funding of RH during a one-month stay at INLN (Nice) where, among other projects,  
this work was conceived. Finally, RH would like to thank Thierry Grandou of INLN for his hospitality 
and very useful conversations.       








\appendix

\onecolumn 

\section{Code adjustments in \texttt{recfast}}

In Appendix A we exhibit the modifications of code \texttt{recfast} due to SU(2)$_{\rm CMB}$. 

\begin{center}
	\setlength{\LTcapwidth}{\textwidth}
	\begin{longtable}{ll}
\caption{Differences in \texttt{recfast} code of \protect\cite{recfastBarc2015} for $\Lambda$CDM versus SU(2)$_{\text{CMB}}$. For a given code line (first column) the first (second) line in second column corresponds to $\Lambda$CDM (SU(2)$_{\text{CMB}}$).\label{codetable}} \\
		\toprule
		line  & recfast  \\
		\midrule
		\endfirsthead
		
		\multicolumn{2}{c}%
		{{\tablename\ \thetable{} -- continued from previous page}} \\
		\hline \multicolumn{1}{l}{line} &
		\multicolumn{1}{l}{recfast} \\ \hline 
		\endhead
		
		\hline \multicolumn{2}{r}{{Continued on next page}} \\ 
		\endfoot
		
		\bottomrule
		\endlastfoot
		
		356 &\footnotesize\verb|fnu = (21.d0/8.d0)*(4.d0/11.d0)**(4.d0/3.d0)| \\
		 &  \footnotesize\verb|fnu = (21.d0/8.d0)*(16.d0/23.d0)**(4.d0/3.d0)| \\
		358 & \footnotesize\verb|z_eq = (3.d0*(HO*C)**2/(8.d0*Pi*G*a*(1.d0+fnu)*Tnow**4))*OmegaT|  \\
		&\footnotesize\verb|z_eq = (3.d0*(HO*C)**2/(8.d0*Pi*G*a*(4.0d0+fnu)*(Tnow*0.63d0)**4))*OmegaT| \\
		421 & \footnotesize\verb|y(3) = Tnow*(1.d0+z)| \\
		& \footnotesize\verb|y(3) = Tnow*(1.d0+z)*0.63d0| \\
		462 & \footnotesize\verb|if (zend.gt.8000.d0) then| \\
		 & \footnotesize\verb|if (zend.gt.13000.d0) then| \\
		 469  & \footnotesize\verb|y(3) = Tnow*(1.d0+z)| \\
		 & \footnotesize\verb|y(3) = Tnow*(1.d0+z)*0.63d0| \\
		 471 & \footnotesize\verb|else if(z.gt.5000.d0)then| \\
		 & \footnotesize\verb|else if(z.gt.8000.d0)then| \\
		 475 & \footnotesize\verb|rhs = dexp( 1.5d0 * dLog(CR*Tnow/(1.d0+z))| \\
		 & \footnotesize\verb|rhs = dexp( 1.5d0 * dLog(CR*Tnow*0.63d0/(1.d0+z))| \\
		 476 & \footnotesize\verb|- CB1_He2/(Tnow*(1.d0+z)) ) / Nnow| \\
		 & \footnotesize\verb|- CB1_He2/(Tnow*(1.d0+z)*0.63d0) ) / Nnow| \\
		 482 & \footnotesize\verb|y(3) = Tnow*(1.d0+z)| \\
		 & \footnotesize\verb|y(3) = Tnow*(1.d0+z)*0.63d0| \\
		 484 & \footnotesize\verb|else if(z.gt.3500.d0)then| \\
		 & \footnotesize\verb|else if(z.gt.5650.d0)then| \\
		 491 & \footnotesize\verb|y(3) = Tnow*(1.d0+z)| \\
		 & \footnotesize\verb|y(3) = Tnow*(1.d0+z)*0.63d0| \\
		 496 & \footnotesize\verb|rhs = dexp( 1.5d0 * dLog(CR*Tnow/(1.d0+z))| \\
		 & \footnotesize\verb|rhs = dexp( 1.5d0 * dLog(CR*Tnow*0.63d0/(1.d0+z))| \\
		 497 & \footnotesize\verb|- CB1_He1/(Tnow*(1.d0+z)) ) / Nnow| \\
		 & \footnotesize\verb|- CB1_He1/(Tnow*0.63d0*(1.d0+z)) ) / Nnow| \\
		 505 & \footnotesize\verb|y(3) = Tnow*(1.d0+z)| \\
		 & \footnotesize\verb|y(3) = Tnow*(1.d0+z)*0.63d0| \\
		 509 & \footnotesize\verb|rhs = dexp( 1.5d0 * dLog(CR*Tnow/(1.d0+z))| \\
		 & \footnotesize\verb|rhs = dexp( 1.5d0 * dLog(CR*Tnow*0.63d0/(1.d0+z))| \\	
		 510 & \footnotesize\verb|- CB1/(Tnow*(1.d0+z)) ) / Nnow| \\
		 & \footnotesize\verb|- CB1/(Tnow*0.63d0*(1.d0+z)) ) / Nnow| \\
		 525 & \footnotesize\verb|Trad = Tnow * (1.d0+zend)| \\
		 & \footnotesize\verb|Trad = Tnow * (1.d0+zend)*0.63d0| \\
		 560 & \footnotesize\verb|if(z.gt.8000.d0)then| \\
		 & \footnotesize\verb|if(z.gt.13000.d0)then| \\
		 566 & \footnotesize\verb|else if(z.gt.3500.d0)then| \\
		 & \footnotesize\verb|else if(z.gt.5650.d0)then| \\
		 570 & \footnotesize\verb|rhs = dexp( 1.5d0 * dLog(CR*Tnow/(1.d0+z))| \\
		 & \footnotesize\verb|rhs = dexp( 1.5d0 * dLog(CR*Tnow*0.63d0/(1.d0+z))| \\
		 571 & \footnotesize\verb|- CB1_He2/(Tnow*(1.d0+z)) ) / Nnow| \\
		 & \footnotesize\verb|- CB1_He2/(Tnow*0.63d0*(1.d0+z)) ) / Nnow| \\
		 576 & \footnotesize\verb|else if(z.gt.2000.d0)then| \\
		 & \footnotesize\verb|else if(z.gt.3200.d0)then| \\
		 579 & \footnotesize\verb|rhs = dexp( 1.5d0 * dLog(CR*Tnow/(1.d0+z))| \\
		 & \footnotesize\verb|rhs = dexp( 1.5d0 * dLog(CR*Tnow*0.63d0/(1.d0+z))| \\
		 580 & \footnotesize\verb|- CB1_He1/(Tnow*(1.d0+z)) ) / Nnow| \\
		 & \footnotesize\verb|- CB1_He1/(Tnow*0.63d0*(1.d0+z)) ) / Nnow| \\
		 589 & \footnotesize\verb|rhs = dexp( 1.5d0 * dLog(CR*Tnow/(1.d0+z))| \\
		 & \footnotesize\verb|rhs = dexp( 1.5d0 * dLog(CR*Tnow*0.63d0/(1.d0+z))| \\
		 590 & \footnotesize\verb|- CB1/(Tnow*(1.d0+z)) ) / Nnow| \\
		 & \footnotesize\verb|- CB1/(Tnow*0.63d0*(1.d0+z)) ) / Nnow| \\
		 660 & \footnotesize\verb|Trad = Tnow * (1.d0+z)| \\
		 & \footnotesize\verb|Trad = Tnow*0.63d0 * (1.d0+z)| \\
		 805 & \footnotesize\verb|f(3) = Tnow| \\
		 & \footnotesize\verb|f(3) = Tnow*0.63d0| \\
	\end{longtable}
\end{center}

\section{Fit of the low-$z$ behaviour of $T(z)$ in SU(2)$_{\rm CMB}$}

\begin{figure}
\centering
\includegraphics[width=0.7\columnwidth]{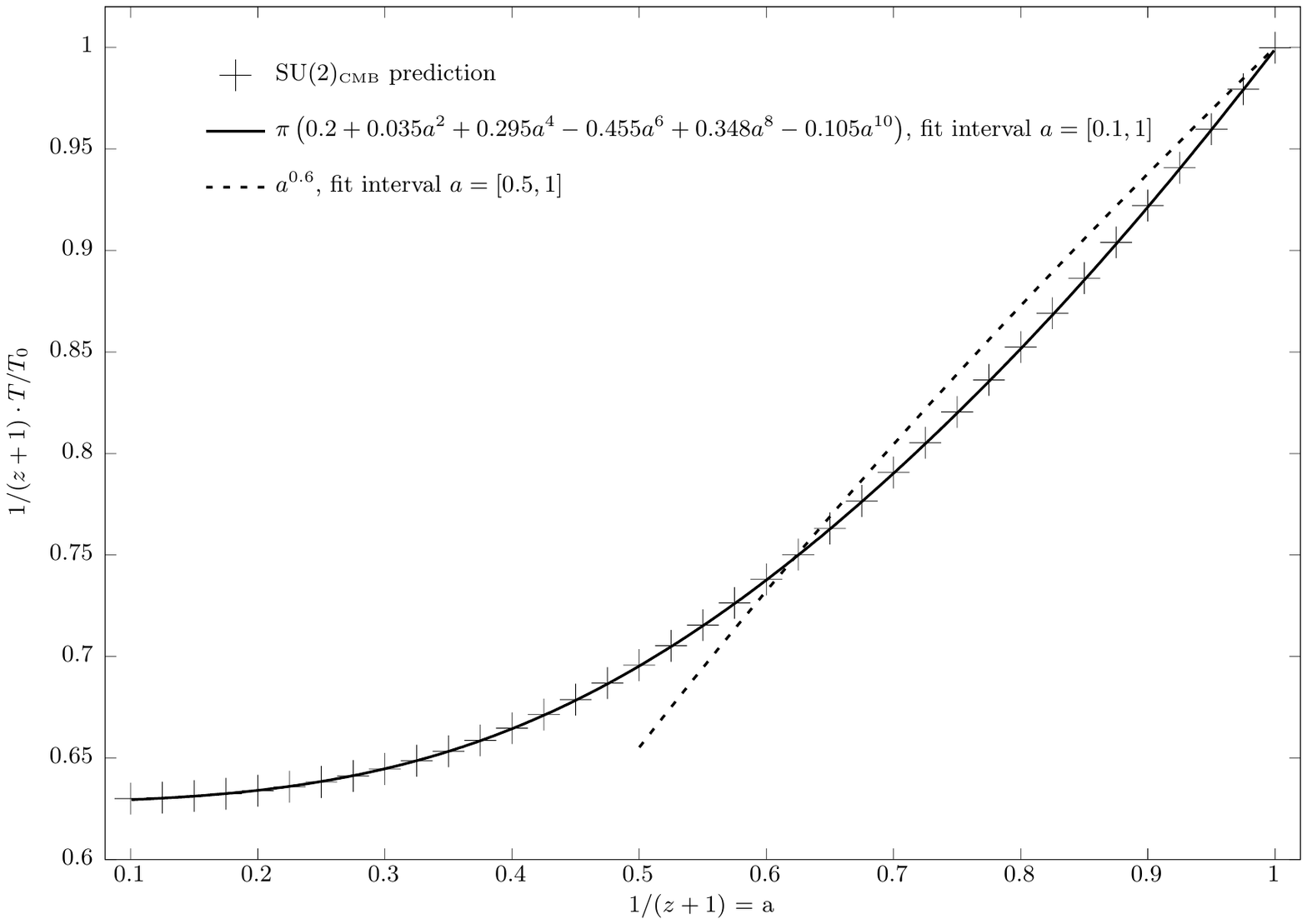}
\caption{Low-$z$ behaviour of function $\frac{1}{z+1}\frac{T}{T_0}$ for SU(2)$_{\rm CMB}$. Crosses 
denote the prediction from SU(2)$_{\rm CMB}$, the solid line represents the 
best fit to an even-power polynomial of degree ten in $\frac{1}{z+1}$ for $0\le z\le 9$, and the 
dashed line shows the best fit to the power law $(z+1)^{-\beta}$ for $0\le z\le 1$ ($\beta=0.6$). \label{Fig-B1}}   
\end{figure}

In Appendix B we show that $T(z)$, obtained by solving the equation for energy conservation of 
SU(2)$_{\rm CMB}$ in an FLRW universe \cite{Hofmann2015} (with a slight and inessential 
correction of the high-$z$ coefficient from 0.62 to 0.63), cannot be well 
fitted to the power law $T(z)/T_0=(1+z)^{1-\beta}$ ($\beta$ constant) 
assumed in "extractions" of $T(z)$ from the thermal Sunyaev-Zel'dovich effect for $0\le z\le 1$, see, e.g.  
\cite{Luzzi}. This is demonstrated in Fig.\,\ref{Fig-B1}

\section{Baryon velocity decoupling}

Appendix C provides arguments why the commonly used criterion involving the drag depth $\tau_{\rm drag}$, 
as introduced in \cite{Hu1996} to characterise the freeze-out of baryon 
velocity $v_b$ during recombination, is imprecise. Namely, we show that condition 
$\tau_{\rm drag}=1$ essentially determines the maximum $z_{\rm max,drag}$ of a localized, yet finite-width 
distribution $D_{\rm drag}(z^\prime,z)$ (with variable $z^\prime$ and, essentially, independence of $z$) during 
recombination. It is the $z^\prime$-integral over 
$D_{\rm drag}(z^\prime,z)$ down to $z$ which determines the freeze-out behaviour of 
$v_b$ in $z$. Here freeze-out of 
$v_b$ means $v_b\propto a^{-1}$. Thus $v_b a$ can be 
considered constant in $z$ only if the $z^\prime$-integration starts to include the left flank (lf) 
of $D_{\rm drag}(z^\prime,z)$, that is for $z\le z_{\rm lf,drag}$.  
For a (positive) bump-like distribution the left-flank position 
$z_{\rm lf,drag}$ coincides with the position $z_{\rm max,drag}$ of the maximum of 
$\frac{dD_{\rm drag}}{dz^\prime}$. As we shall show below, $z_{\rm lf,drag}$ is considerably lower than 
$z_{\rm max,drag}$ for both SU(2)$_{\rm CMB}$ and $\Lambda$CDM.  

\begin{figure}
\centering
\subfloat{\includegraphics{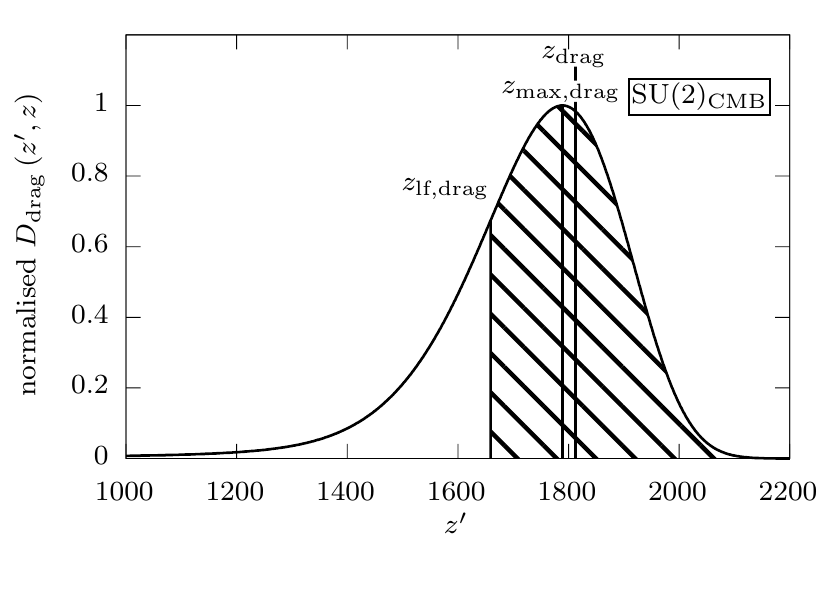}}
\subfloat{\includegraphics{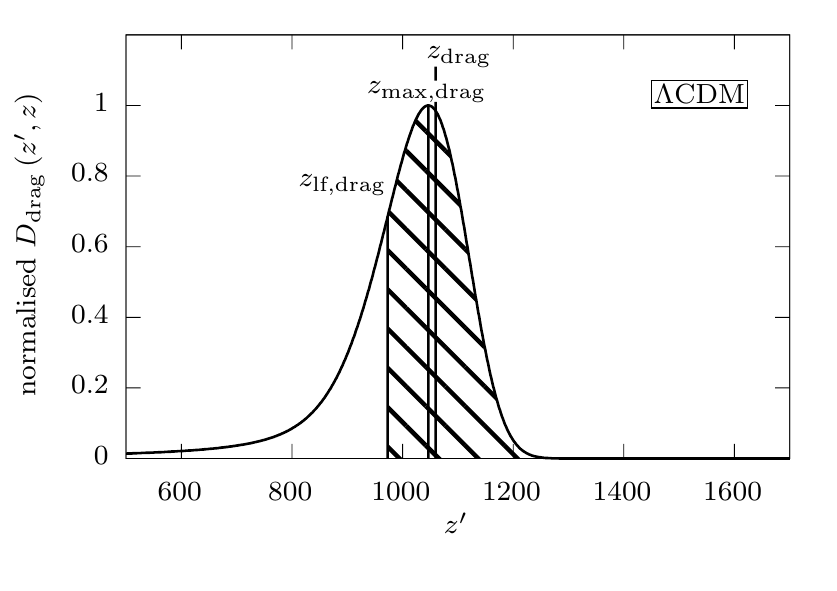}}
\caption{Normalised function $D_{\rm drag}(z^\prime,z)$, defined in Eq.\,(\ref{Ddef}), 
if $z\le z_{\rm max,drag}$ for SU(2)$_{\rm CMB}$ (left) and $\Lambda$CDM (right). Redshift $z_{\rm lf,drag}$ is defined 
as the position of the maximum of $\frac{dD_{\rm drag}}{dz^\prime}$ (position of left flank of 
$D_{\rm drag}$) whereas $z_{\rm max,drag}$ denotes the position of the maximum of 
$D_{\rm drag}$. The value of $z_{\rm drag}$, defined in Eq.\,(\ref{zdragdef}), essentially 
coincides with $z_{\rm max,drag}$: $z_{\rm drag}=1813\sim z_{\rm max,drag}=1789$ for SU(2)$_{\rm CMB}$ 
and $z_{\rm drag}=1059\sim z_{\rm max,drag}=1046$ for $\Lambda$CDM. This should be contrasted 
with $z_{\rm lf,drag}=1659$ for SU(2)$_{\rm CMB}$ and $z_{\rm lf,drag}=973$ for $\Lambda$CDM. The hatched area under the curve determines the freeze-out value of $a v_b$.   \protect{\label{Fig-C1}}.}     
\end{figure}

\begin{figure}
\centering
\subfloat{\includegraphics{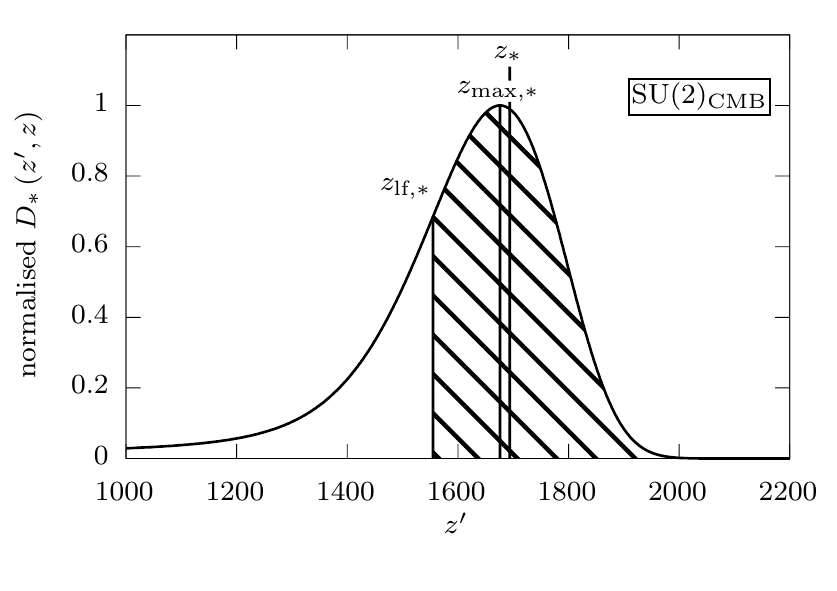}}
\subfloat{\includegraphics{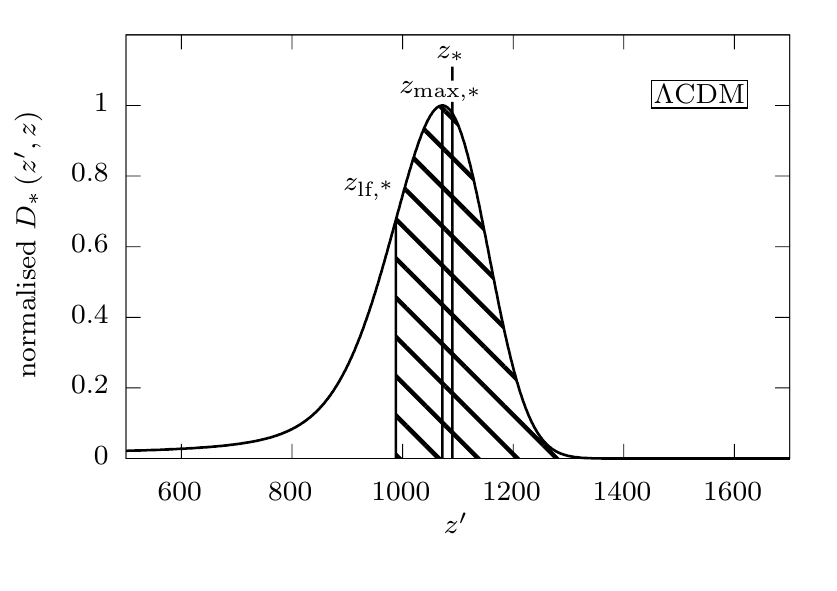}}
\caption{Normalised function $D_{*}(z^\prime,z)$ in analogy to Eq.\,(\ref{Ddef}) but now for photon decoupling, 
if $z\le z_{{\rm max},*}$ for SU(2)$_{\rm CMB}$ (left) and $\Lambda$CDM (right). Redshift $z_{{\rm lf},*}$ is defined 
as the position of the maximum of $\frac{dD_{*}}{dz^\prime}$ (position of left flank of 
$D_{*}$) whereas $z_{{\rm max},*}$ denotes the position of the maximum of 
$D_{*}$. The value of $z_{*}$, defined in Eq.\,(\ref{optdepthThomson}), essentially 
coincides with $z_{{\rm max},*}$: $z_{*}=1694\sim z_{{\rm max},*}=1676$ for SU(2)$_{\rm CMB}$ 
and $z_{*}=1090\sim z_{{\rm max},*}=1072$ for $\Lambda$CDM. This should be contrasted 
with $z_{{\rm lf},*}=1555$ for SU(2)$_{\rm CMB}$ and $z_{{\rm lf},*}=988$ for $\Lambda$CDM. 
The hatched area under the curve determines the freeze-out value of the temperature perturbation.\protect{\label{Fig-C2}}}     
\end{figure}

The decoupling of the CMB photons and baryon velocity at co-moving wave number $k$ (omitted as a subscript in the following)
during recombination is described by a Boltzmann hierarchy for the temperature perturbation $\Theta_l$ \cite{BondEfstathiou1984} \footnote{By virtue of \cite{PeeblesWilk} 
$\Theta_1$ can be interpreted as the velocity of the photon fluid.}, the Einstein-Poisson equations for the metric and Newtonian gravitational potential, $\Phi$ and $\Psi$, respectively, and the continuity and Euler 
equations for the coupled baryon-photon fluid \cite{Hu1994}. Only the Euler equation is 
needed for the following argument. It reads \cite{Hu1994}
\eqb
\label{Euler}
\dot{v}_b=-\frac{\dot{a}}{a}v_b+k\Psi+\dot{\tau}_{\rm drag}(\Theta_1-v_b)\,, 
\eqe
where overdots represent derivatives w.r.t. conformal time,
$\eta=\int_0^t \frac{dt^\prime}{a(t^\prime)}$, $\dot{\tau}_{\rm drag}\equiv\frac{\dot{\tau}}{R}$ with 
$\dot{\tau}\equiv\chi_e n^b_e\sigma_T a$ and $R$, $n^b_e$, $\chi_e$ defined in Eqs.\,(\ref{Rdef}), (\ref{neb}), (\ref{iondef}), respectively. As usual, $\sigma_T$ denotes the Thomson cross section. Varying the coefficient $K$ in the solution
\eqb
\label{homEuler}
av_b=K\e^{-\int_0^\eta d\eta^\prime\,\dot{\tau}_{\rm drag}(\eta^\prime)}\equiv K\e^{-\tau_{\rm drag}(\eta)}
\eqe
of the homogeneous part 
\eqb
\label{Eulerhomeq}
\dot{v}_b=-\frac{\dot{a}}{a}v_b-\dot{\tau}_{\rm drag}v_b 
\eqe
of Eq.\,(\ref{Euler}), we obtain the following solution to the full equation (\ref{Euler})
\eqb
\label{fullEuler}
av_b(\eta)=\lim_{\epsilon\searrow 0}\int_{\epsilon}^\eta d\eta^\prime\,\e^{-\tau_{\rm drag}(\eta^\prime,\eta)}a(\eta^\prime)\left(\dot{\tau}_{\rm drag}(\eta^\prime)\Theta_1(\eta^\prime)+k\Psi(\eta^\prime)\right)\,,
\eqe
where $\tau_{\rm drag}(\eta^\prime,\eta)\equiv \int_{\eta^\prime}^\eta d\eta^{\prime\prime}\,\dot{\tau}_{\rm drag}(\eta^{\prime\prime})$, and $av_b$ is subject to the (Big-Bang) initial condition $a(0)=0$. 
Ignoring the effect of the gravitational potential\footnote{Since there is no cold dark matter in SU(2)$_{\rm CMB}$ potential wells become important only long after recombination when the dark-matter component of the late-time $\Lambda$CDM model is present, see Appendix D.} $\Psi$, the authors of \cite{Hu1996} argue that, independently of $\eta$, the factor 
\eqb
\label{facF}
F_{\rm drag}(\eta^\prime,\eta)\equiv \e^{-\tau_{\rm drag}(\eta^\prime,\eta)}
\dot{\tau}_{\rm drag}(\eta^\prime)
\eqe
behaves like a delta function, centered at $\eta^\prime=\eta_{\rm max}$, which would imply 
that decoupling\footnote{Since we here consider co-moving 
distances on the scale of the sound horizon only it is justified to assume that $\Theta_1$ is a slowly varying function.} 
occurs for $\eta\ge \eta_{\rm max}$. This, however, is imprecise since $F_{\rm drag}$ has a finite 
width. Namely, re-writing the solution (\ref{fullEuler}) in terms of redshift $z$, we have 
\begin{align}
\label{FullEulerz}
av_b(z)&=\lim_{Z\nearrow \infty}\int_{z}^{Z} dz^\prime\,\frac{\e^{-\tau_{\rm drag}(z^\prime,z)}}{H(z^\prime) (z^\prime+1)}\left(\dot{\tau}_{\rm drag}(z^\prime)\Theta_1(z^\prime)+k\Psi(z^\prime)\right)\nonumber\\ 
&\sim\lim_{Z\nearrow \infty}\int_{z}^{Z} dz^\prime\,D_{\rm drag}(z^\prime,z)\Theta_1(z^\prime)\,,
\end{align}
where (with a slight abuse of notation)
\eqb
\label{tauz}
\tau_{\rm drag}(z^\prime,z)\equiv \int_{z}^{z^\prime}dz^{\prime\prime}\,\frac{\dot{\tau}_{\rm drag}(z^{\prime\prime})}{H(z^{\prime\prime})}\,.
\eqe
In Eq.\,(\ref{FullEulerz}) we have defined  
\eqb
\label{Ddef}
D_{\rm drag}(z^\prime,z)\equiv \frac{\e^{-\tau_{\rm drag}(z^\prime,z)}
\dot{\tau}_{\rm drag}(z^\prime)}{H(z^\prime) (z^\prime+1)}\,,
\eqe 
and, for the reason given above, the gravitational potential $\Psi$ was ignored in going from the first to the second line. 
Fig.\,\ref{Fig-C1} indicates $D_{\rm drag}$ as a function of $z^\prime$ ($z$-independence) 
for SU(2)$_{\rm CMB}$ and $\Lambda$CDM. Notice the closeness of $z_{\rm drag}$ and 
$z_{\rm max,drag}$ in both models. In contrast, $z_{\rm lf,drag}$ turns out to be considerably lower.  
It is worth mentioning that condition (\ref{optdepthThomson}), again, essentially describes the position 
of the maximum $z_{{\rm max},*}$ of the according function $D_{*}(z^\prime,z)$ appearing in the formal, approximate 
solution of the Boltzmann hierarchy for the temperature perturbation \cite{Hu1996}, $z_{*}\sim z_{{\rm max},*}$.
Also here $D_{*}(z^\prime,z)$ is broad, and one 
should use $z_{{\rm lf},*}$ instead of $z_{*}$ as a more 
realistic redshift for photon decoupling, see Fig.\,\ref{Fig-C2}.     

\section{(De-)Percolating Planck-scale axionic solitons and interpolation of high-$z$ with low-$z$ cosmology}

Appendix D proposes a cosmological model to interpolate low-$z$ $\Lambda$CDM with high-$z$ 
SU(2)$_{\rm CMB}$. The basic idea invokes the fact that a Planck-scale axion field 
$\varphi$ (a pseudo Nambu-Goldstone field of dynamical chiral symmetry breaking \cite{Adler1969,AdlerBardeen1969,Bell1969,Fujikawa1979,Fujikawa1980} near the Planck 
scale \cite{Neubert}) due to non-thermal phase transitions of the Hagedorn type in the early 
universe forms U(1) topological solitons (vortices) subject to a size (and mass) distribution characterised by 
several distinct peaks. Since SU(2)$_{\rm CMB}$ is the only deconfining-phase Yang-Mills 
theory up to temperatures reaching far beyond recombination, $\varphi$'s potential is given as 
\cite{Peccei1977a,Peccei1977b} 
\begin{equation}
\label{poten}
V\left(\varphi \right) = \left(\kappa \Lambda_\text{CMB}\right)^4 \cdot \left(1-\cos\left(\varphi/m_\text{P}\right)\right)\,,
\end{equation}
where $\Lambda_{\rm CMB}\sim 10^{-4}\,$eV, $\kappa$ is a dimensionless factor of order unity, the 
reduced Planck mass reads 
\eqb
\label{defplamass}
m_{\rm P}\equiv \frac{1.22\times 10^{19}}{\sqrt{8\pi}}\,\mbox{GeV}=(8\pi G)^{-1/2}\,,
\eqe
and $G$ denotes Newton's constant. Complemented by a canonical kinetic term and 
assuming minimal coupling to gravity, Eq.\,(\ref{poten}) is the 
basis for the derivation of the according field equations to describe the 
self-gravitating vortex-like solitons.  

Increasing $z$, the density of such solitons within a given, highly populated  
narrow size-band reaches a critical value at $z_p$ where percolation into 
homogeneous and time-independent energy density occurs. (We assume instantaneous percolation). 
Assuming that only one such percolation point $z_p$ 
occurs within $z=0$ and $z_{p^\prime}\gg z_{{\rm lf}, *}$, we have  
 \begin{equation}
\label{Haxion}
H^2 = \frac{8 \pi G}{3} \left(\rho_b + \rho_{\rm DS } + \rho_r \right)
\equiv \frac{8 \pi G}{3}\rho_c\,.
\end{equation}
Here $\rho_r$ denotes radiation-like energy density 
including SU(2)$_{\rm CMB}$ and 
three flavours of massless neutrinos\footnote{For $z\le 9$ radiation energy density is 
severely suppressed in the cosmological 
model, for $z>9$ the thermal ground state and the masses of 
the vector modes of SU(2)$_{\rm CMB}$ can be neglected.}. In Eq.\,(\ref{Haxion}) we may approximate $\rho_r$ as 
\begin{equation}
\label{radne}
\rho_r = \Omega_{\gamma,0}\rho_{c,0} \cdot \left\{ \begin{array}{cl}0 &\quad  (z < 9)  \\ 
A\left(1+\frac{7}{32}\left(\frac{16}{23}\right)^{4/3}\,N_{\nu}\right)\left(z+1\right)^4 & \quad (z \geq 9)
\end{array}\right.\,, 
\end{equation}
with $A=4(0.63)^3$, $N_{\nu}=3$, and $\Omega_{\gamma,0}=4.6\times 10^{-5}$, compare with Eq.\,(\ref{Hzexp}). Furthermore,  $\rho_b=\Omega_{b,0}\rho_{c,0}(z+1)^3$ is the energy density of 
baryons. We set $\Omega_{b,0}=0.04$. Finally, $\rho_{\rm DS}$ represents the 
dark-sector energy density, representing free vortices (dark matter) or percolated vortices (dark energy), given as  
\begin{equation}
\label{DSed} 
\rho_{\rm DS }=\Omega_\Lambda\rho_{c,0}+\Omega_{{\rm DM},0}\rho_{c,0}\cdot \left\{ \begin{array}{cl}
\left(z_{\phantom{p}} +1\right)^3 & \quad (z<z_p) \\
\left(z_p+1\right)^3 & \quad (z\geq z_p)
\end{array}  \right.\,, 
\end{equation}
where $\Omega_{{\rm DM},0}\rho_{c,0}$ is today's pressureless dark-sector energy 
density, represented by a gas of Planck-scale-axion vortices, and 
$\Omega_\Lambda\rho_{c,0}$ denotes constant vacuum energy associated 
with yet percolated Planck-scale-axion vortices. 
We set $\rho_{c,0}=\frac{3} {8 \pi G}H_0^2$ with 
$H_0=73.24$\,km\,s$^{-1}$\,Mpc$^{-1}$ \cite{Riess2016}, and we use 
$\Omega_{{\rm DM},0}=0.26$ and $\Omega_\Lambda=0.7$.  

The observable angular scale $\theta_*$ 
of the sound horizon $r_s$ at the redshift $z_{{\rm lf},*}$ of CMB photon decoupling 
is given as
\begin{equation}
\label{angscale}
\theta_*=\frac{r_s(z_{{\rm lf},*})}{\int_0^{z_{{\rm lf},*}}\frac{dz}{H(z)}}\,. 
\end{equation}
To match $\theta_*=0.597^\circ$ fitted in \cite{Ade2016} we require $z_p=155.4$. This yields a percentage of vacuum energy at CMB 
photon decoupling of about 
\eqb
\label{vaceneatPD}
\frac{\Omega_{{\rm DM},0}}{\Omega_{b,0}}\left(\frac{z_p+1}{z_{{\rm lf},*}+1}\right)^3\sim 0.65\%\,.
\eqe
The omission of vacuum energy in our SU(2)$_{\rm CMB}$ high-$z$ cosmological 
model of Eq.\,(\ref{Hzexp}) thus is justified for the interpolating model 
proposed in Appendix D. 
\begin{figure}
\centering
\includegraphics[width=0.6\columnwidth]{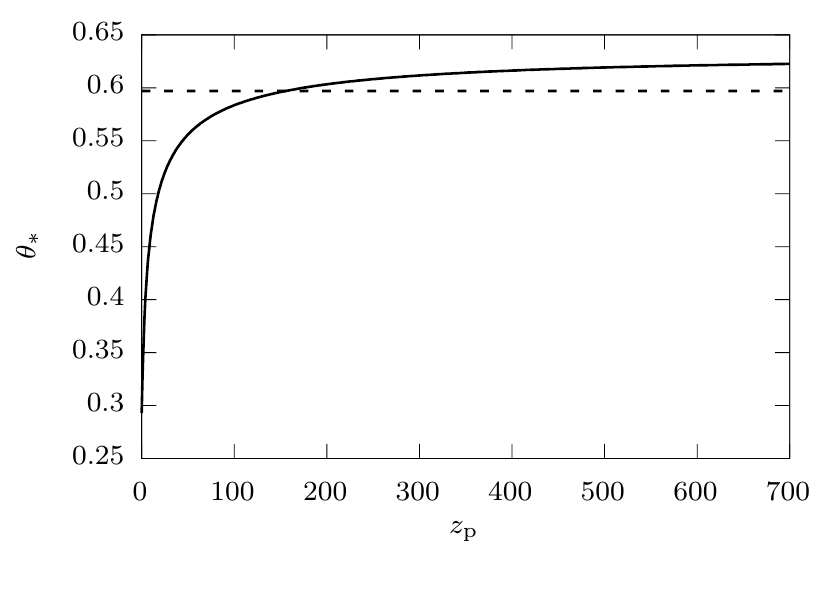}
\caption{Function $\theta_*(z_p)$ for $\Omega_{\Lambda}=0.7$, $\Omega_{{\rm DM},0}=0.26$, 
$\Omega_{b,0}=0.04$, $\Omega_{\gamma,0}=4.6\times 10^{-5}$, and 
$H_0=73.24$\,km\,s$^{-1}$\,Mpc$^{-1}$ for the high-$z$ SU(2)$_{\rm CMB}$ and low-$z$ $\Lambda$CDM interpolating cosmological 
model considered in this Appendix. Also indicated is the value $\theta_*=0.59^\circ$ (dashed line), fitted to the CMB TT power spectrum.} 
\label{Fig-D1}    
\end{figure}

\bsp	
\label{lastpage}
\end{document}